\documentclass[11pt]{article}

 \usepackage{setspace}
\usepackage{amssymb}
\usepackage{amsmath}
\usepackage{epsfig}

\begin{document}
\newtheorem{theorem}{Theorem}
\newtheorem{corollary}{Corollary}
\newtheorem{definition}{Definition}
\newtheorem{lemma}{Lemma}

\newcommand{\define}{\stackrel{\triangle}{=}}

\def\QED{\mbox{\rule[0pt]{1.5ex}{1.5ex}}}
\def\proof{\noindent\hspace{2em}{\it Proof: }}

\date{}
\title{\Large Elements of Cellular Blind Interference Alignment --- \\
Aligned Frequency Reuse, Wireless Index Coding and Interference Diversity} 
\author{Syed A. Jafar\vspace{0.1cm}\\
{\small Center for Pervasive Communications and Computing}\\
{\small Department of Electrical Engineering and Computer Science}\\
{\small University of California Irvine, Irvine, California, 92697}\\
       }
\maketitle
   \abstract

We explore degrees of freedom (DoF) characterizations of partially connected wireless networks, especially cellular networks, with no channel state information at the transmitters. Specifically, we introduce three fundamental elements --- aligned frequency reuse, wireless index coding and interference diversity --- through a series of examples, focusing first on infinite regular arrays, then on finite clusters with arbitrary connectivity and message sets, and finally on heterogeneous settings with asymmetric multiple antenna configurations.  Aligned frequency reuse refers to the optimality of orthogonal resource allocations in many cases, but according to unconventional reuse patterns  that are guided by interference alignment principles. Wireless index coding highlights both the intimate connection between the index coding problem and cellular blind interference alignment, as well as the added complexity inherent to wireless settings. Interference diversity refers to the observation that in a wireless network each receiver experiences a different set of interferers, and depending on the actions of its own set of interferers, the interference-free signal space at each receiver fluctuates differently from other receivers, creating opportunities for robust applications of blind interference alignment principles. 

\thispagestyle{empty}
\newpage
\section{Introduction}

We use the compact phrase \emph{Cellular Blind Interference Alignment} (CB) to  refer broadly to the study of degrees of freedom (mainly interference alignment) of partially connected (mainly cellular) wireless networks with no channel state information (blind) at the transmitters. Since the class of partially connected wireless networks is quite extensive, our goal is not to be exhaustive but to shed light into the non-trivial nature of the problem, and highlight the basic principles  relevant to these settings through carefully chosen examples. We start with a motivating example, and then summarize the background for this work.

\subsection{Locally Connected 4-Cell Downlink Network}

\begin{figure}[!h]
\centering
\includegraphics[width=5.1in]{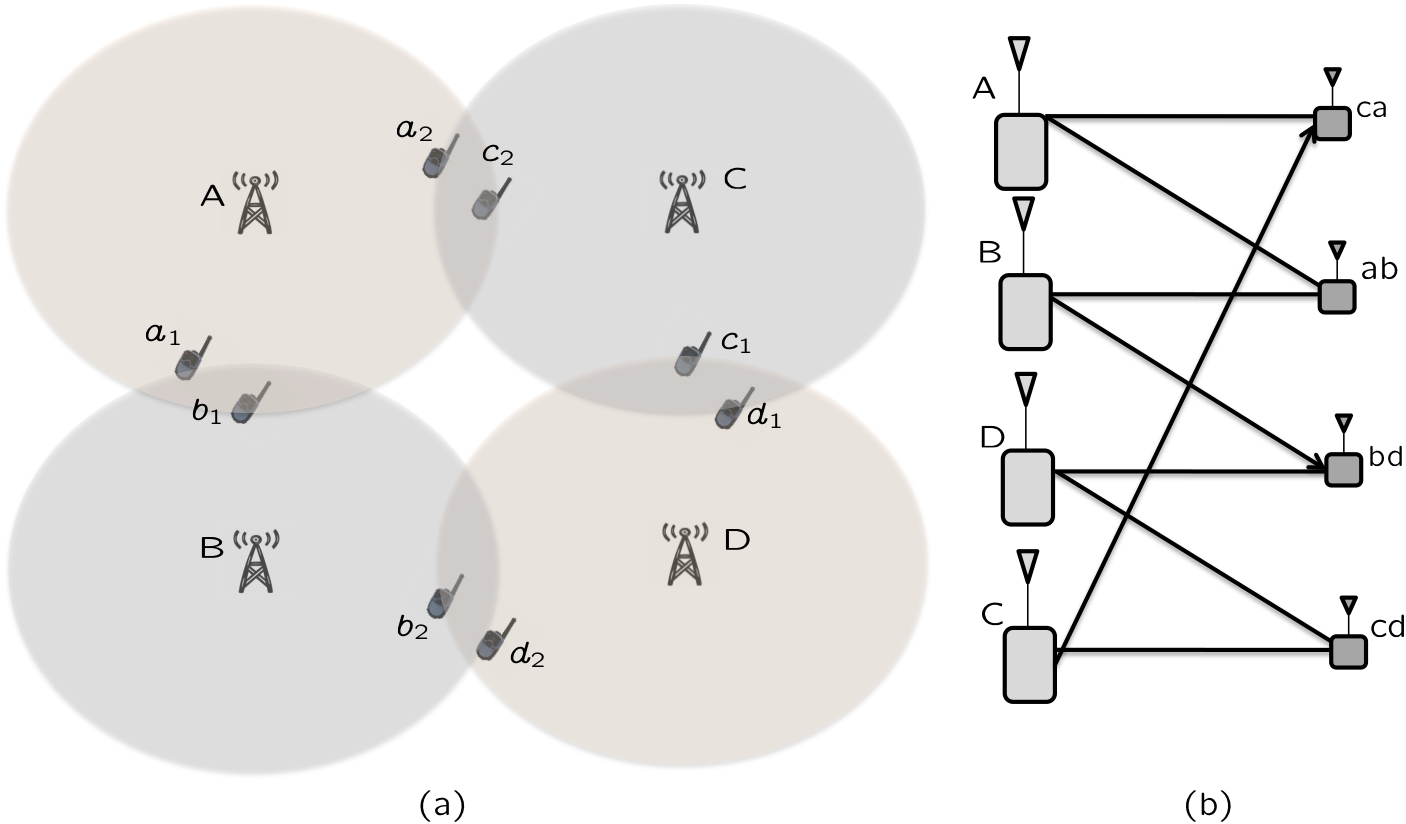}
\caption{\small\it (a) Locally connected 4-cell downlink network, (b) Equivalently, a partially connected X network}
\label{fig:intro}
\end{figure}

Consider the 4 cell downlink setting shown in Fig. \ref{fig:intro}(a). To model propagation path loss the network is assumed to be partially (locally) connected. Specifically, each base station is heard only within the cell area shown as a shaded disk centered at that base station. This gives rise to interference for the users located in the overlapping cellular regions. In the figure we show two users in each boundary region, one for each of the two cells that overlap at that boundary. For example, receivers $a_1, a_2$ want to receive independent messages $W_{a1}, W_{a2}$ from the base station $A$, but have to deal with interference from base stations $B, C$, respectively. 

In order to emphasize robustness, suppose we wish to investigate this partially connected network under fairly pessimistic assumptions ---  that all the non-zero channels are independent and identically distributed (i.i.d.) at any given time, and vary in an i.i.d. manner across blocks according to the standard block fading model applied identically to all users, there is only one antenna at each transmitter and receiver, there is no channel state information at the transmitters (no CSIT) other than the knowledge of the topology (connectivity) of the network, and the receivers only know their own channels from the base stations that they can hear.  Because the users inhabiting the same boundary are statistically indistinguishable from the transmitters' perspective, they have the same ability to decode messages and hence may be combined into one equivalent user without loss of generality, as shown in Fig. \ref{fig:intro}(b), producing a partially connected X network setting \cite{Cadambe_Jafar_X} with 8 independent messages, one from each transmitter to each connected receiver. So, e.g., the receiver labeled `$ab$' wants to decode  two independent messages sent from transmitters labeled $A, B$. 

We wish to find out the degrees of freedom (DoF) of this setting.
A natural scheme would be  an orthogonal (frequency reuse) scheme which eliminates interference, e.g., cells $A, D$ are simultaneously active for half the time  (or over half the frequency band) and cells $C, D$ are simultaneously active for the remaining half of the time. This orthogonal scheme achieves 0.5 DoF per cell, and it is easy to see that no orthogonal scheme can achieve higher DoF.  But \emph{ is this the optimal DoF value for this network? } Because there is no CSIT, one might think so. The reader might enjoy giving this question some thought to find the best possible achievable DoF value  before proceeding to the solution which will be presented in a later section of this paper.

\subsection{Background}
The idea of interference alignment   highlights both the opportunities and the challenges inherent in a wireless communication network \cite{Jafar_FnT}. On the one hand, interference alignment schemes show remarkable potential to increase the number of available signal dimensions \cite{Cadambe_Jafar_int}. On the other hand, most interference alignment schemes assume  \emph{global} availability of vast amounts of channel knowledge (even if it is delayed \cite{Maddah_Tse, Maleki_Jafar_Shamai}), which implies excessive training  and feedback overheads. The extent to which these substantial overheads are essential, is the central issue that could  define the role of interference alignment schemes in future wireless networks. One particularly interesting aspect of this broader question is the feasibility of interference alignment without any knowledge (not even delayed knowledge) of the  actual channel coefficient realizations at the transmitters, and only \emph{local} channel knowledge at the receivers.

The challenging nature of the problem is evident in the observation that the degrees of freedom  with no channel state information at the transmitter(s), i.e., \emph{DoF with no CSIT, remain unknown in general} for all those wireless networks where interference alignment is relevant in the presence of CSIT. Conventional wisdom, consistent with a few conjectures \cite{Lapidoth_Shamai_Wigger_BC, Weingarten_Shamai_Kramer}, has consistently predicted a collapse of DoF, and therefore the infeasibility of interference alignment, with no CSIT. However, except under additional assumptions of homogeneity\footnote{It is important to clarify that while i.i.d. isotropic fading is the most commonly studied example of the ``no CSIT" setting, the ``no CSIT" assumption  does not necessarily  imply isotropic fading conditions or statistical equivalence of receivers. Knowledge of long term channel statistics, e.g., the fading distribution, coherence intervals etc., which can vary from user to user,  may be assumed within the no CSIT framework. What is not allowed in the no CSIT setting is the knowledge of the actual channel coefficient realizations.} such as isotropic fading conditions, degradedness or statistical indistinguishability of receivers where indeed the collapse of DoF has been established \cite{Huang_Jafar_Shamai_Vishwanath, Zhu_Guo_MIMOIC, Guo_isotropic, Varanasi_noCSIT}, the DoF remain unknown. A case in point is the Lapidoth-Shamai-Wigger conjecture \cite{Lapidoth_Shamai_Wigger_BC} on the collapse of DoF of a MISO BC (multiple input single output broadcast channel) which remains unresolved, except when the users are (essentially) statistically indistinguishable. On the other hand, there are recent results that support the feasibility of interference alignment even with channel uncertainty. A notable example is the compound setting with arbitrary (finite) number of states where the DoF are shown to be robust to channel uncertainty \cite{Gou_Jafar_Wang}.  

Perhaps the strongest evidence so far, of the feasibility of  interference alignment with no CSIT, is presented in \cite{Jafar_corr, Wang_Gou_Jafar, Wang_Gou_Jafar_MIMO} through the idea of \emph{blind} interference alignment. The central idea of {blind} interference alignment, introduced in \cite{Jafar_corr}, is that the knowledge of distinct channel coherence patterns associated with different receivers, without any knowledge of the actual channel coefficient realizations, may be enough to achieve interference alignment. For example, the MISO BC with two antennas at the base station, where one user experiences primarily time-selective fading and the other user experiences primarily frequency-selective fading, is shown to achieve $\frac{4}{3}$ DoF (which is also the information theoretic DoF outer bound for this setting), without any knowledge of the actual channel coefficient realizations at the transmitter. Since the overhead of learning long term channel statistics (in this case the channel coherence intervals) is small, this is easily one of the  most overhead-free examples of interference alignment. However, there is a caveat here, in that one must rely on nature to create suitably heterogeneous coherence intervals for blind interference alignment. Another apparent limitation is that the scheme relies strongly on the block fading model, requiring both that the channel coefficients stay  constant within a coherence interval and that they  change from one block to another. While the requirements can be clearly softened for practical applications at finite SNR by limiting to those settings where the channel coefficients are \emph{sufficiently} constant over a coherence interval and \emph{sufficiently} distinct across coherence intervals, it does considerably narrow down the scope of practical applications. 

An alternative application of blind interference alignment is introduced in \cite{Wang_Gou_Jafar, Wang_Gou_Jafar_MIMO} where instead of relying on nature to provide suitable heterogeneous coherence intervals, distinct antenna switching patterns are employed at various receivers that are equipped with reconfigurable antennas (equivalently, antenna selection). Assuming that the length of the switching pattern is well within the coherence interval of all channels, small scale coherence patterns are artificially imposed by the switching of antennas.  While no feedback overhead is required to communicate the coherence patterns resulting from antenna switching (which is performed in a pre-determined fashion), the caveat here is the need for reconfigurable antennas, which limits the application of this form of blind interference alignment.

\subsection{Robustness Assumptions}
In this work we continue the search for more robust blind interference alignment principles. As a measure of progress, at the very least we wish to avoid the limitations of the previous approaches. We would like to neither be dependent on nature to create coherence patterns that vary from user to user in a suitable manner, nor require additional hardware in the form of reconfigurable antennas to artificially impose those patterns. In order to allow only robust schemes, we  pre-emptively enforce the following 5  assumptions throughout this work.

\begin{enumerate}
\item {\it Ordinary Coherence Model: } Much of this work, especially the next section on aligned frequency reuse requires no assumptions on physical coherence intervals.  In subsequent sections, when we  do require physical channel coherence,  we ensure robustness by applying the standard block fading model  \emph{identically} to all channels, so that all channels have the same coherence time, called the network coherence time. Note that the blind alignment solution of \cite{Jafar_corr}, which requires naturally occurring heterogeneous block fading models, is precluded. 
\item {\it Ordinary antennas:} The antennas are not reconfigurable, i.e., there exists no mechanism to artificially switch the channel coefficient value between two nodes. Thus, the blind alignment solution of \cite{Wang_Gou_Jafar, Wang_Gou_Jafar_MIMO} does not apply.
\item {\it Blind Transmitters:} The transmitters have no knowledge of actual channel coefficient values. Thus there is no feedback overhead for acquiring CSIT.
\item {\it Local CSIR:} Channel state information at the receiver (CSIR) is limited to only the channel coefficients directly associated with the receiver. Thus, each channel coefficient value is precisely known to only one node (the associated receiver), and there is no overhead of sharing channel coefficients among nodes. 
\item {\it Partial Connectivity:}  While we assume that the transmitters do not know the actual channel coefficient realizations,  a knowledge of the \emph{connectivity} of the network, is assumed. 
\end{enumerate}

It is the last assumption, of local  (or partial) connectivity, that is the most critical\footnote{What is critical is the significant difference in signal strengths from adjacent base stations versus those that are farther away. It is notable that the Wyner type connectivity model, where distant base stations are assumed to be entirely disconnected, is used in this work only for analytical simplicity and is not a make-or-break assumption. Extensions to GDoF settings for example can be made to include more elaborate connectivity models.}.  Wireless networks  invariably possess a locally connected character due to the universality of propagation path loss. The main contribution of this work is the idea that local or partial connectivity naturally creates opportunities for robust interference alignment with no CSIT. Since local connectivity is most directly associated with cellular networks, we refer to this research direction as \emph{cellular blind interference alignment}.

\section{The Cellular Blind Interference Alignment Problem $\mathcal{CB}(\mathcal{C},\mathcal{W}_t, \mathcal{W}_r)$}
\subsection{Problem Formulation}
Let $\mathcal{T}=\{t_1,t_2,\cdots \}$ be the index set of transmitters, $\mathcal{R}=\{r_1,r_2,\cdots\}$ be the index set of receivers, and $\mathcal{W}=\{W_1, W_2,\cdots\}$  be the set of messages in a partially connected wireless network.  The number of elements in these sets corresponds to the number of transmitters, receivers and messages, respectively, and may be infinite for extended networks. All messages are independent. 

A  cellular blind interference alignment setting is defined  by a connectivity set $\mathcal{C}$ and a mapping of messages to transmitters and receivers.
\begin{eqnarray}
\mathcal{C}&\define&\{(r,t) ~\mbox{ such that } c_{rt}=1, r\in\mathcal{R}, t\in\mathcal{T}\}
\end{eqnarray}
where $c_{rt}\in\{0,1\}, r\in\mathcal{R}, t\in\mathcal{T}$, is a binary random variable that takes the value $1$ if receiver $r$ is ``connected" to transmitter $t$, and $0$ otherwise. Let us define message sets corresponding to each transmitter and each receiver.
\begin{eqnarray}
\forall ~ t\in\mathcal{T}, &&~\mathcal{W}_t\subset\mathcal{W}\\
\forall~r\in\mathcal{R},&&~\mathcal{W}_r\subset\mathcal{W}\\
\forall~ t\neq t',&& ~\mathcal{W}_t\cap\mathcal{W}_{t'}=\phi\\
&&\bigcup_{t\in\mathcal{T}}\mathcal{W}_t=\mathcal{W}
\end{eqnarray}
i.e., $\mathcal{W}_t$ is the set of messages originating at transmitter $t\in\mathcal{T}$, $\mathcal{W}_r$ is the set of messages desired by receiver $r\in\mathcal{R}$,  each message originates  at a unique transmitter but may be intended for multiple receivers.

With this notation, the locally connected 4-cell network presented in the introduction is described as follows.
\begin{eqnarray}
\mathcal{T}&=&\{A, B, C, D\}\\
\mathcal{R}&=&\{a_1, a_2, b_1, b_2, c_1, c_2, d_1, d_2\}\\
\mathcal{W}&=&\{a_1, a_2, b_1, b_2, c_1, c_2, d_1, d_2\}
\end{eqnarray}
\begin{eqnarray*}
&&\mathcal{W}_A=\{a_1,a_2\}, \mathcal{W}_B=\{b_1,b_2\},\mathcal{W}_C=\{c_1,c_2\},\mathcal{W}_D=\{d_1,d_2\}\\
&&\mathcal{W}_{a_1}=\{a_1\}, \mathcal{W}_{a_2}=\{a_2\}, \mathcal{W}_{b_1}=\{b_1\}, \mathcal{W}_{b_2}=\{b_2\}, \mathcal{W}_{c_1}=\{c_1\}, \mathcal{W}_{c_2}=\{c_2\}, \mathcal{W}_{d_1}=\{d_1\}, \mathcal{W}_{d_2}=\{d_2\}
\end{eqnarray*}
Evidently, detailed notation can be quite cumbersome even for small networks. For this reason, whenever possible, we will simply describe the problem through a network graph. The notation will be useful especially when dealing with arbitrary settings, e.g., in Section \ref{sec:arbitrary} where  we will refer to arbitrary problem formulations as $\mathcal{CB}(\mathcal{C},\mathcal{W}_t, \mathcal{W}_r)$.

Further context is provided by the channel model.
\subsection{Channel Model}
While we will adapt the notation from section to section to keep it from becoming too cumbersome,  the basic channel model throughout this work is the  standard Gaussian network, described as:
\begin{eqnarray}
\forall r\in\mathcal{R}, ~~~Y_{r}(n) &=& \sum_{t\in\mathcal{T}} c_{rt}H_{rt}(n)X_t(n)+Z_r(n)
\end{eqnarray}
where during the $n^{th}$ channel use,  $Y_r(n)$ is the observed signal at receiver $r$, $Z_r(n)$ is the additive white Gaussian noise term, $H_{rt}(n)$ is the channel coefficient between transmitter $t$ and receiver $r$,  $X_t(n)$ is the  symbol sent from transmitter $t$, and $c_{rt}\in\{0,1\}$ is the connectivity parameter. Note that connectivities are assumed to be fixed for the duration of communication.   The input symbols are subject to a power constraint $P$, i.e., $\forall t\in\mathcal{T}$, E$|X_{t}|^2\leq P$.  We will assume all symbols are complex. The AWGN is assumed to be zero mean unit variance circularly symmetric complex Gaussian. To avoid degeneracies, each channel coefficient magnitude is restricted to a finite interval bounded away from zero and infinity. The receivers  have knowledge of their channels from all transmitters  to which they are connected, i.e., receiver $r$ knows $H_{rt}(n)$ for all $t$ such that $c_{rt}=1$. We refer to this as the \emph{local CSIR} assumption, because no receiver is  required to learn channels associated with other receivers.

Next we describe the assumptions regarding channel random processes $\left[H_{rt}(n)\right]$ across space ($r,t$) and across time ($n$). In the spatial dimension, we will always assume channel variables are independent across $r,t$. If we additionally impose the condition that $\left[H_{rt}(n)\right]$ are identically distributed across $r,t$ then we simply refer to it as the \emph{spatially i.i.d. fading model}. 

In the time dimension, we make the block fading assumption where all  channel coefficients are constant within a network coherence block and switch to an independent realization from one block to the next. The fading across blocks is assumed i.i.d. according to a continuous distribution. There are some subtle issues with block fading models,  that we overcome through a novel set of assumptions, which we explain next. 

 The block fading model captures the channel coherence phenomenon which is certainly realistic. However, a predetermined setting of block boundaries gives rise to artificial ``edge effects" because the achievable schemes, which may require symbol extensions, must conform to the externally imposed block structure. To avoid these artificial issues, and to retain the focus on essential aspects, we will assume that \emph{the choice of the channel fading block length, $\tau$, is in the hands of the codebook designer}, up to a finite pre-set upper bound, $\tau_{\max}$ and subject to the constraint that all channels have the same coherence block length. In other words, the physical network coherence time is captured by the parameter $\tau_{\max}$ which is intended to be a coarse metric, as coherence time should be in practice. Within the confines of this coarse coherence interval, an achievable scheme can decide precisely how much coherence it needs without worrying about edge effects by \emph{setting} the theoretical network coherence interval $\tau$ to any integer value in the range $[0, \tau_{\max}]$, which must be held fixed throughout the duration of communication. A converse must therefore hold for all possible choices of coherence intervals in this range. Note that $\tau$, chosen offline, is a constant network parameter known to all nodes. In general we will only distinguish between two cases: the assumption that $\tau_{\max}$ is large enough  will simply be called  ``sufficient coherence" in time, and the assumption that  $\tau_{\max}=1$ will be called i.i.d. fading in time.
 
To summarize, we  have four basic  statistical channel models: i.i.d. or sufficient coherence in time, and i.i.d. or not (only independent but not identically distributed) in space. Note that the temporal statistics do not distinguish one channel from another, only spatial statistics (if not assumed i.i.d.) can.

Regarding CSIT, the transmitters are aware only of the channel statistics, i.e., properties that do not change with $n$. This includes the connectivities $c_{rt}$, and the network coherence time $\tau$, and of course, does \emph{not} include channel coefficients $H_{rt}(n)$.

The small set of basic models is chosen for convenience rather than necessity. Most of our results are valid for less restrictive assumptions than the chosen models. Occasionally, we will highlight the generalizations. For instance, in the next section, all the results continue to hold as long as the  channels have magnitudes bounded away from zero and infinity. Beyond this, we need no other assumption. The channels may or may not be block fading, may be drawn from continuous distributions or chosen arbitrarily, may not even  be ergodic or stationary, they may stay constant or vary, independently or according to some arbitrary spatial and temporal correlation model, it will not impact the DoF results.

Finally, the achievable rates, capacity region and degrees of freedom are defined in the standard Shannon theoretic sense, and we are interested in the DoF value for the network.



\section{Aligned Frequency Reuse}

\begin{figure}[!h]
\centering
\includegraphics[width=4.2in]{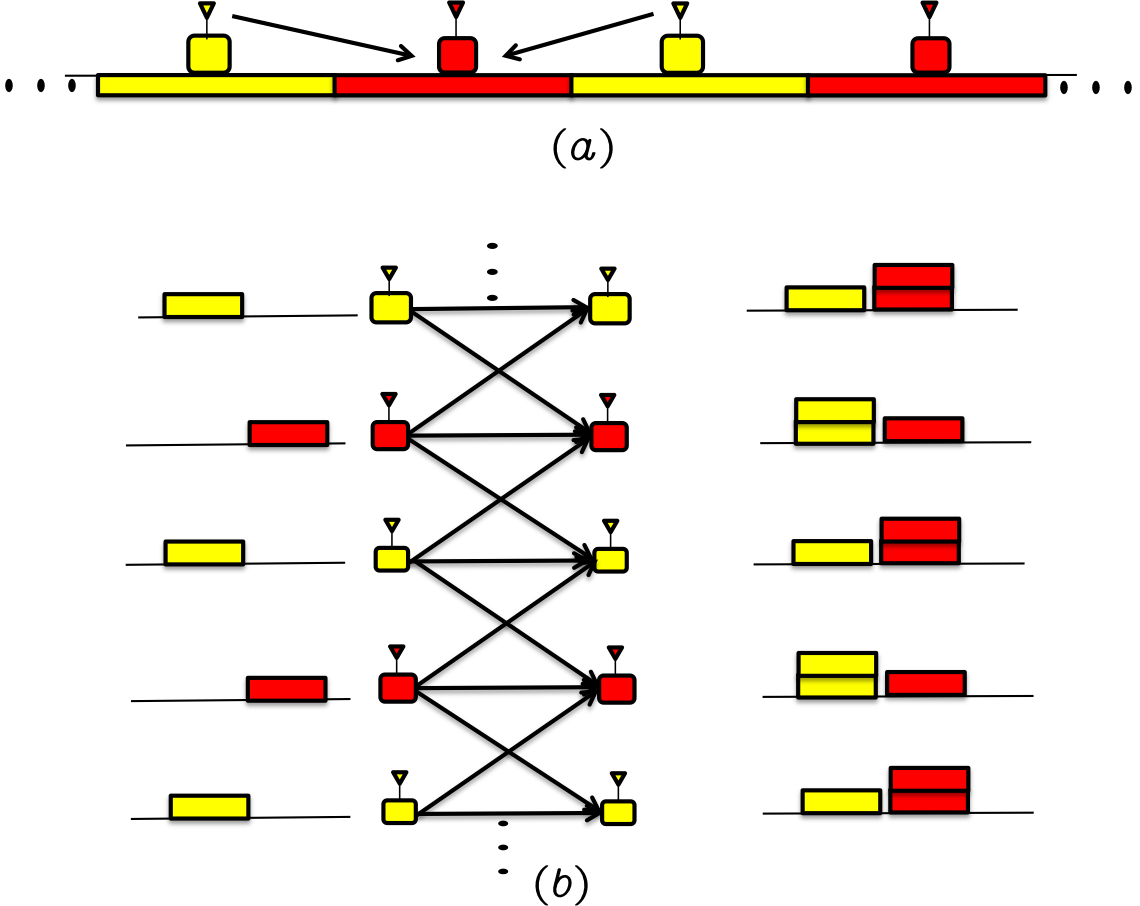}
\caption{\it Frequency reuse as a simple form of interference alignment. (a) A simple frequency reuse pattern, and (b) An interference alignment setting. Red and yellow represent two halves of the spectrum. Alignment is achieved at the receivers as the two undesired interferers occupy the same half of the spectrum while the desired signal occupies the other half.}
\label{fig:reuse}
\end{figure}

Frequency reuse, i.e., an orthogonal allocation of signaling dimensions based on connectivity, is a central concept for cellular networks. For example, consider the simplified cellular downlink setting shown in Fig. \ref{fig:reuse}(a) where a series of base stations is evenly spaced along a straight line, defining an infinite array of cells. Due to path loss, each base station is heard within its own cell and to a certain extent in its two neighboring cells, but no further than that. The conventional frequency reuse plan for this setting is to divide the total spectrum into two equal and orthogonal sub-bands and alternately allocate these bands to successive cells. Note that all inter-cell interference is avoided by this simple scheme. Here we make an interesting observation --- that this \emph{frequency-reuse  is in fact a simple form of interference alignment}. Each cell experiences two interferers besides its own desired signal, and yet each cell is able to access one-half of the total spectrum. This is accomplished only because both interferers are aligned within the same half of the spectrum.  The similarity is explicitly evident from Fig. \ref{fig:reuse}(b), where the analogous (locally connected) $K$ user interference channel setting is shown.  Moreover, since no knowledge of channel coefficient values is need at the transmitters, this simple example is in fact a (trivial) case of \emph{blind interference alignment}.

The interpretation of cellular frequency-reuse as interference alignment offers us a new perspective. It also highlights the significance of path-loss in the study of IA schemes. $K/2$ DoF are achieved (and also optimal) in the $K$ user interference channel, regardless of whether it is fully connected (no path loss) or locally connected (due to path loss). The difference in the two settings is in the degree of difficulty. Without path loss one would typically need highly sophisticated IA schemes, requiring instantaneous, global, perfect channel knowledge. However, with path loss, i.e., local connectivity, achieving $K/2$ DoF is as simple as the alternating orthogonal frequency re-use pattern illustrated in Fig. \ref{fig:reuse}.

\begin{figure}[!h]
\centering
\includegraphics[width=6in]{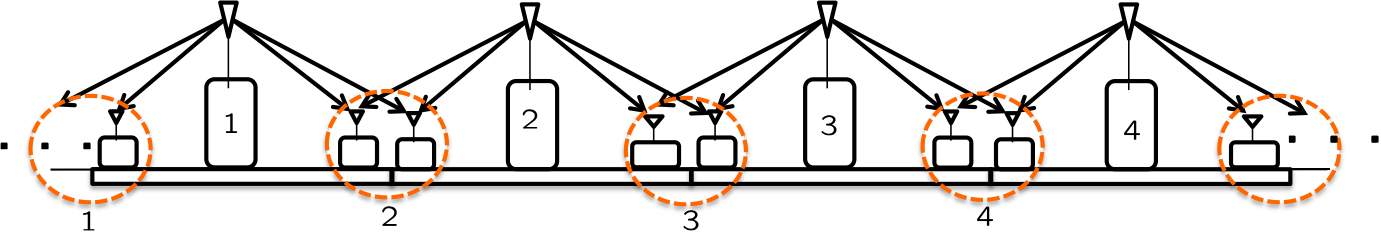}
\caption{\it Locally connected one-dimensional cellular array where boundary users hear both adjacent base stations.}
\label{fig:linear}
\end{figure}

In this section we  consider  infinite arrays of uniformly placed cells that are locally connected. Three topologies are considered: 1) Linear Cellular Array, where cells are placed uniformly along a straight line, 2) Square Cellular Array, where the cell placements fall on a square grid, and 3) Hexagonal Cellular Array, where the cells are placed uniformly in a hexagonal grid pattern.  We are interested primarily in the users located at the boundaries between adjacent cells, since this is where interference is the most severe and where frequency planning is most needed.\footnote{This is consistent with the fractional frequency reuse principle used, e.g., in Mobile WiMAX,  which allows users near the cell center to follow a reuse factor of 1 and requires frequency planning only for cell edge users.}  A one-dimensional cellular model is shown in Fig. \ref{fig:linear}. Since in a linear model, each cell has 2 adjacent cells with which it shares boundaries, Fig. \ref{fig:linear} shows two representative users in each cell, one at each shared boundary.

\subsection{Linear Cellular Array}
\begin{figure}[!h]
\centering
\includegraphics[width=6in]{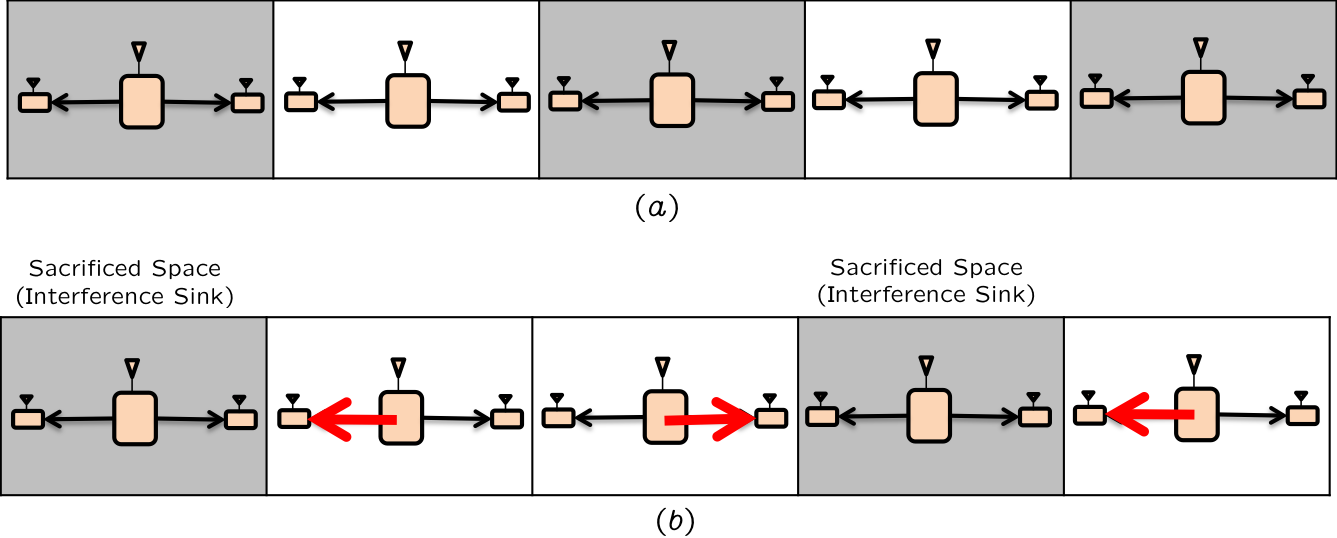}
\caption{\small\it (a) Conventional frequency reuse (periodic with period 2) allows 1/2 DoF per cell. (b) Aligned frequency reuse (periodic with period 3) allows 2/3 DoF per cell, an improvement of 33\% per cell. All transmissions are isotropic. Red arrows point out active receivers.}
\label{fig:linearsol}
\end{figure}

As shown in Fig. \ref{fig:linear}, each base station is heard by all users around the cell boundary, both in its own cell as well as the immediately adjacent cells. However, due to path loss, the signals do not travel further beyond. Conventional spectral-reuse pattern used in this setting is shown in Fig. \ref{fig:linearsol}(a) where active cells that share a common cell edge are assigned different spectral bands, or equivalently, only alternating cells are activated, the even numbered cells for half the time, and the odd numbered cells for the remaining half of the time. Thus, each cell achieves 1/2 DoF and all inter-cell interference is eliminated. { The conventional frequency reuse solution, 1/2 DoF per cell, is the baseline} that we compare against, as we present next an alternative approach ---- aligned frequency reuse.

Aligned frequency reuse is illustrated in Fig. \ref{fig:linearsol}(b), where a periodic reuse pattern with period 3 cells is repeated along the infinite sequence of cells. Every third cell (shown as the grey shaded cells in Fig. \ref{fig:linearsol}(b)) is sacrificed, i.e., switched off. Since the transmitter of the sacrificed cell generates no signal, all the boundaries of sacrificed cell become interference free and therefore all the neighboring cells can serve users that are located on the boundary of the sacrificed cell\footnote{Note that the red arrows in Fig. \ref{fig:linearsol} do not indicate directional transmission (all transmission is isotropic), but rather the choice of the receiver to be served within each active cell. The remaining receivers are turned off.}. Clearly the throughput per cell and per user can be symmetrized by shifting the pattern so that each cell becomes the sink cell for 1/3 of the time. The resulting bandwidth allocation from this \emph{aligned} frequency reuse is 2/3 DoF per cell, which corresponds to { an improvement of $33\%$ over the baseline}. Next we establish the optimality of this aligned frequency reuse pattern.

\begin{theorem}
For the locally connected linear cellular model illustrated in Fig. \ref{fig:linear}, the optimal DoF value is 2/3 DoF per cell. 
\end{theorem}
\proof
First let us prove the information theoretic DoF outer bound. Recall that each base station has two independent messages, one for each user located on either boundary. Because of the symmetry of the problem, there is no loss of generality in assuming that each message carries the same DoF. Now we claim that each message cannot have more than 1/3 DoF.  Let us consider any user in Fig. \ref{fig:linear}, e.g, the user in Cell 2 located at the boundary with Cell 3. Let us eliminate all messages except the desired message from BS 2 and the two undesired messages from BS 3. Clearly eliminating other messages cannot hurt the rates of the remaining messages. Now we argue that all three remaining messages are resolvable by this user. Since the desired message is decodable by design, the user can reliably reconstruct and subtract it from its received signal. This gives the user an invertible channel to BS 3 (within bounded variance noise distortion), from which it can reliably resolve both messages originating at BS 3. Thus, one user is able to resolve all 3 messages. Since the user has only 1 antenna, the total DoF of all three messages cannot be more than 1. This gives us an outer bound of $1/3$ DoF per message. Since each cell has two messages, it gives us an outer bound of $2/3$ DoF per cell. 

Achievability of 2/3 DoF per cell is already shown in Fig. \ref{fig:linearsol}(b), and clearly requires no knowledge of channel realizations at the transmitters. \hfill\QED

{\it Remark: } It is remarkable that not only is the 2/3 DoF per cell the optimal DoF value in the absence of channel knowledge at the transmitters, but also that it is the optimal value \emph{even with global and instantaneous channel knowledge}. The difference between the two settings is that if we consider multiple users at each boundary, the DoF outer bound without CSIT will still be only 2/3 per cell (assuming here the users are statistically equivalent in the absence of CSIT), but with global CSIT the DoF per cell will approach 1 as the number of users increases, i.e., as if all interference is eliminated.

\subsection{Square Cellular Array}
The one-dimensional setting shows that the benefits of aligned frequency reuse can be significant, i.e., 33\% improvement over the  baseline corresponding to conventional frequency reuse. As we go from one dimensions to two, perhaps the most interesting question is to determine if the benefits diminish, increase, or remain the same.

\begin{figure}[!h]
\centering
\includegraphics[width=6in]{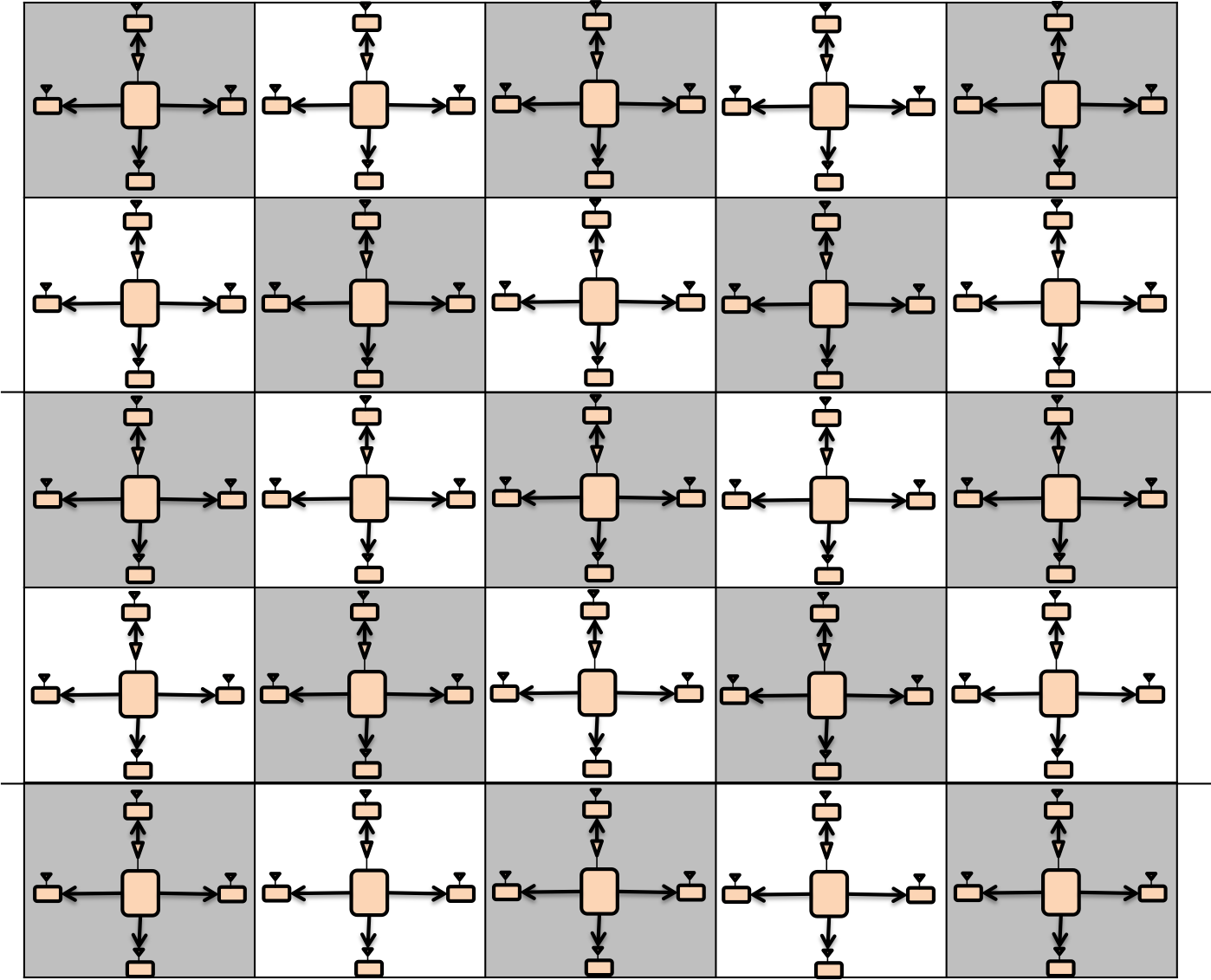}
\caption{\small\it Baseline frequency reuse pattern, achieves 1/2 DoF per cell.}
\label{fig:baselinearray}
\end{figure}

To answer this question, we consider a two dimensional square cellular grid with the same local connectivity assumptions as in the previous section. We assume, as before, that each user on the boundary between two cells, hears comparable signal strengths from both base stations, and each base station can be heard within its own cell and in the vicinity of its boundary with its adjacent cells. The conventional frequency reuse solution for this setting is shown in Fig. \ref{fig:baselinearray}, and corresponds to 1/2 DoF per cell as before.

\begin{figure}[!h]
\centering
\includegraphics[width=6in]{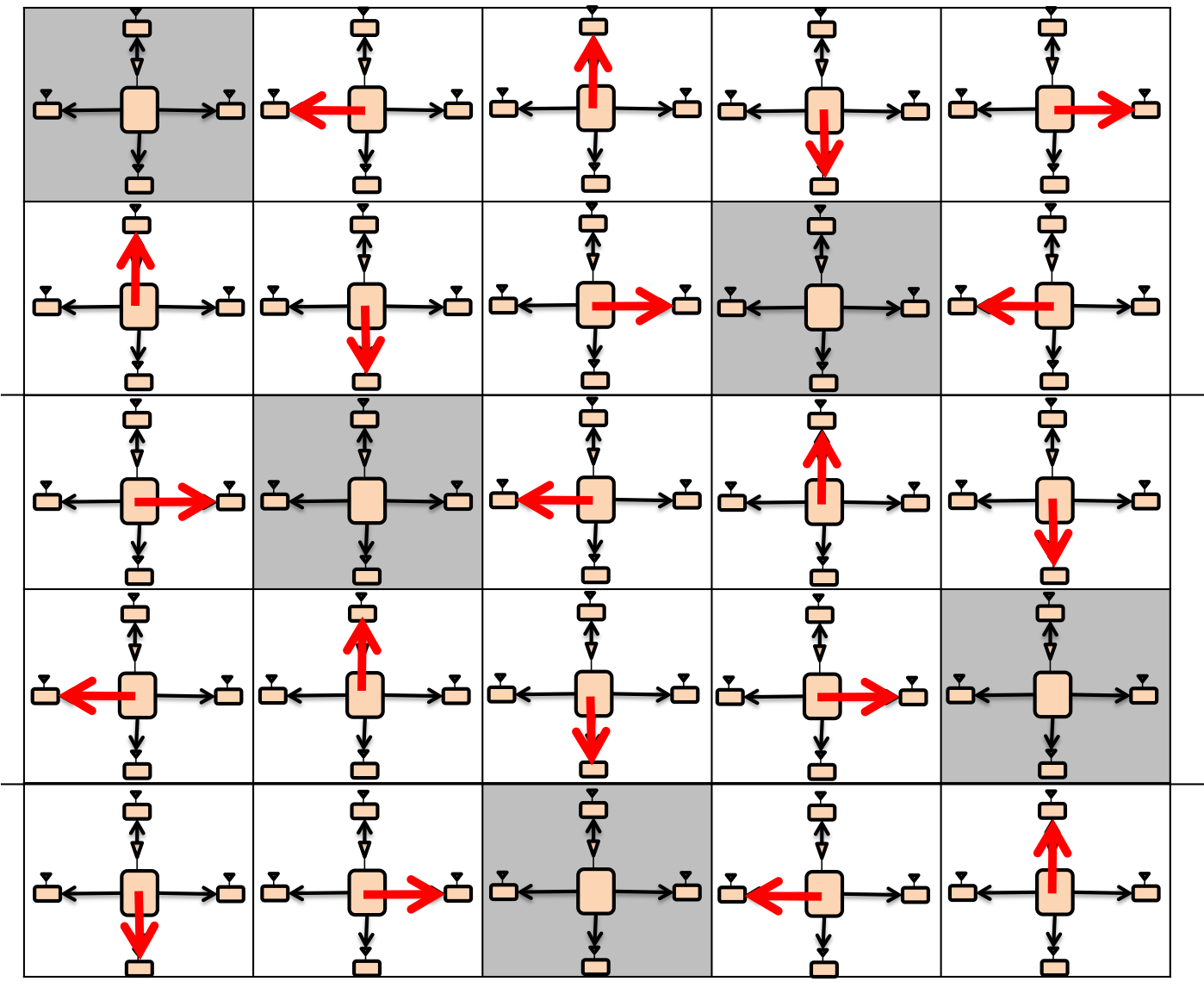}
\caption{\small\it Aligned frequency reuse, 60\% improvement over the baseline. All transmissions are isotropic. Red arrows only point out active receivers. All active receivers are adjacent to inactive (grey) cells.}
\label{fig:blindarray}
\end{figure}

The aligned reuse pattern is shown in Fig. \ref{fig:blindarray}, where again the interference sinks are shown in grey. The red arrows indicate the intended users whose messages are being transmitted. Recall that the BS antennas are isotropic, so the \emph{red arrows do not represent directional transmissions}. From Fig. \ref{fig:blindarray} it is clear that the achieved DoF value is 4/5 per cell. This corresponds to an { improvement of 60\% over the baseline}. Next we establish the optimality of this result.

\begin{theorem}
For two dimensional square cellular grid model with local connectivity, as described above, the optimal DoF value is 4/5 DoF per cell. 
\end{theorem}
\proof
The proof of the outer bound is similar to the one-dimensional setting. Consider any user at the boundary between two cells. We claim that this user can resolve not only its desired message, which it must by design, but  also all 4 messages from its neighboring interfering BS. This is because if we eliminate all other messages, then the user can reconstruct and subtract the signal from its desired BS, and then invert the channel to its interfering BS, thus resolving all 5 messages. Since a user with only 1 antenna can resolve 5 messages, the DoF per message cannot be more than 1/5. Then, since each cell has 4 independent messages, the DoF per cell cannot be more than 4/5.

Achievability of 4/5 DoF per cell is already shown in Fig. \ref{fig:linearsol}(b), and clearly requires no knowledge of channel realizations at the transmitters. \hfill\QED

{\it Remark: } Once again it is remarkable that not only is the 4/5 DoF per cell the optimal DoF value in the absence of channel knowledge at the transmitters, but also that it is the optimal value \emph{even with global and instantaneous channel knowledge}. This is because the outer bound argument presented above is not affected by the presence or absence of CSIT.

\subsection{Hexagonal Cellular Array}
\begin{figure}[!h]
\centering
\includegraphics[width=6in]{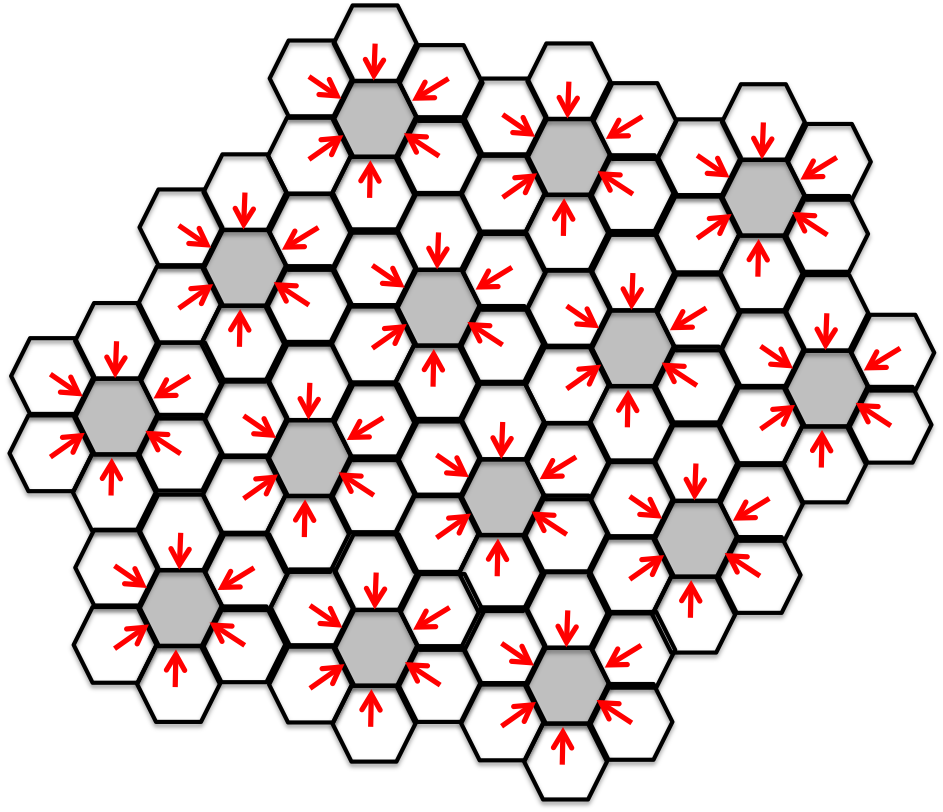}
\caption{\small\it Aligned frequency reuse, 157\% improvement over the baseline. All active receivers are adjacent to inactive cells. While not shown explicitly, users are located on the cell boundaries, one on each side of the boundary. All transmissions are non-directional. Red arrows only indicate the locations of the active users. The base station transmitters in the grey cells are switched off.}
\label{fig:hexarray}
\end{figure}

Next, Fig. \ref{fig:hexarray} shows the aligned frequency reuse pattern for a hexagonal cellular layout. 

\begin{theorem}
For two dimensional hexagonal cellular network with local connectivity, the optimal DoF value is 6/7 DoF per cell. 
\end{theorem}
\proof
The achieved value  is 6/7 DoF per cell, as obvious from the choice of active users illustrated in Fig. \ref{fig:hexarray}. Compared to the conventional 3 frequency reuse pattern, which results in 1/3 DoF per cell, this represents a 157\% improvement per cell. The information theoretic optimality of 6/7 DoF per cell for this setting, both with or without CSIT, follows from similar arguments as the previous theorems. Considering any user let us eliminate all messages except the desired message and all the messages from the interfering base station. These 7 messages cannot have a total of more than 1 DoF because the single user with a single receive antenna can resolve all 7 messages. This follows from the logical argument that the user must be able to reliably decode his desired message by design, so he must be able to subtract the signal from the desired base station, which allows the user to re-construct the transmitted symbol from the interfering base station within bounded variance noise distortion, and thereby decode all 7 messages.\hfill\QED

\subsection{Reciprocal (Uplink) Networks}
While reciprocal networks, i.e., networks obtained by switching the roles of  transmitters and receivers for all messages, typically have the same DoF when global channel knowledge is available, duality is not a property commonly associated with the no CSIT setting. Since we presented all our results in a broadcast channel (BC) or downlink setting, a natural   question is whether the same DoF results hold in the reciprocal (uplink) network. Interestingly, this is the case, as formalized by the following theorem.

\begin{theorem}
The reciprocal networks for the locally connected linear, square and hexagonal cellular networks, have $2/3, 4/5,6/7$ DoF per cell, respectively, both with and without CSIT.
\end{theorem}
In other words, reciprocity holds.\\
\proof Because the DoF optimal achievability schemes presented for the downlink settings are orthogonal, it is easy to see that the same solution can be applied in the reciprocal setting by reversing the direction of communication. The outer bounds also follow from similar principles as before. Consider any base station receiver and eliminate all messages except those desired by this base station and the message from one of the interfering transmitters. After decoding and subtracting out its desired messages, clearly the base station is able to resolve the only remaining signal, from the only interfering transmitter. Thus, if the base station has $N$ neighboring cells with which it shares a boundary, then the total DoF of $N+1$ messages cannot be more than 1. Thus, the DoF per message cannot be more than $1/(N+1)$ and the total DoF per cell cannot be more than $N/(N+1)$. This gives us the outer bound of $2/3, 4/5, 6/7$ DoF per cell for the linear, square and hexagonal settings, respectively.\hfill\QED


While the DoF metric in conjunction with the locally connected models offers new insights, such as the idea of aligned frequency reuse evident in all topologies considered so far, a natural question is the robustness of the results as the connectivity model is enriched to account for reduced but non-zero signal strengths from base stations located farther away. The simplest metric to capture the diversity of signal strengths is the generalized degrees of freedom (GDoF) metric. While in this work we are limited to DoF studies, we note that all our solutions can be just as easily applied (albeit not necessarily optimal) within the GDoF setting by incorporating the signals from farther base stations as the new noise floor. Since the immediately adjacent base stations are already accounted for in our connectivity model, for the direct extension to GDoF setting the only additional parameter that comes into play is the strength of the \emph{next} strongest interfering signal. This is captured by the distance of the receiver to the next nearest base station (not  the immediately adjacent base station). Here, we note that for the linear and square grid models, the distance to the next strongest interferer is the same for both the conventional frequency re-use solution and the aligned frequency re-use solution. Thus both of these settings will maintain their relative advantage over conventional frequency reuse in the GDoF setting. The aligned frequency reuse solution for the hexagonal setting on the other hand, is a bit more sensitive because the next-nearest interferer is  closer than with the conventional frequency re-use setting. While this is a sobering observation, we note that because the DoF improvement (157\%) is much stronger for the hexagonal setting than the linear or square cellular grids (33\% and 60\%, respectively), we do have additional room to accommodate interference. The magnitude of the \emph{net} improvement in the GDoF setting is an intriguing question for future work. Further, in going beyond GDoF, what matters is not only  the strengths of the strongest interferers, but the number of such interferers as well. Since this is only an SNR offset effect it is not visible in DoF or GDoF studies but is quite relevant for performance results at finite SNR. The aligned frequency reuse solution for even the linear and square cellular models will lose some of its projected multiplicative (DoF) gains over conventional frequency reuse at moderate SNR values due to this SNR offset. Finally, we conclude this discussion by reminding the reader that because of the overly pessimistic assumptions regarding CSIT, even modest  gains achieved in this setting bode well for the overall question of practical feasibility of  interference alignment schemes.

\section{Locally Connected 4-Cell  Network}
Having considered only infinite regular arrays of cells so far, we are now interested in finite clusters of cells. While clusters and  irregular cell locations are common even in traditional cellular networks due to the peculiarities of the geographical terrain and propagation environments, they are unavoidable in customer-deployed networks, such as femtocell architectures, which  are increasingly a part of the cellular landscape. Clusters and irregular topologies are also typical for mission-centric or tactical networks. 

In this section, we  focus exclusively on the motivating example  of Fig. \ref{fig:intro} presented at the beginning of this paper, solve it under various assumptions and highlight the  interesting aspects, such as the insufficiency of orthogonal solutions, the role of network coherence time and the amount of channel knowledge needed at the receivers. We  start with the sufficient coherence assumption, and address the issue of orthogonality.

\subsection{Are orthogonal solutions sufficient to achieve optimal DoF ?}
The simplicity of the DoF optimal solutions for the cellular array models studied in the previous section, is quite remarkable. Note that the DoF optimal solutions are \emph{orthogonal} in all DoF results that we have presented so far, in the sense that there is \emph{no interference experienced by any active receiver}, because all undesired transmitters that could be heard by active receivers are switched off. It is natural to wonder if the optimality of orthogonal solutions is  a surprising coincidence or is it somehow a degeneracy forced by our excessively pessimistic assumptions. In other words, \emph{do there always exist  orthogonal solutions that are DoF optimal for partially connected wireless networks with no CSIT, no heterogeneous coherence intervals, no reconfigurable antennas, and only local channel knowledge at each receiver}? 

By solving the motivating example, we show that DoF optimal solutions, even with all the pessimistic assumptions mentioned above, in general cannot be orthogonal. For the motivating example of Fig. \ref{fig:intro} it is clear that orthogonal solutions cannot achieve more than a total of 2 DoF. Here we show that for this network, the optimal DoF value  \emph{with or without} CSIT, is 8/3.

\begin{theorem}\label{thm:4cellsol}
The locally connected 4-cell downlink network shown in Fig. \ref{fig:intro} with sufficient coherence ($\tau_{\max}\geq 3$), has DoF = 8/3, both with or without CSIT, and both with and without the spatially i.i.d. fading assumption.
\end{theorem}
\begin{figure}[!h]
\centering
\includegraphics[width=5in]{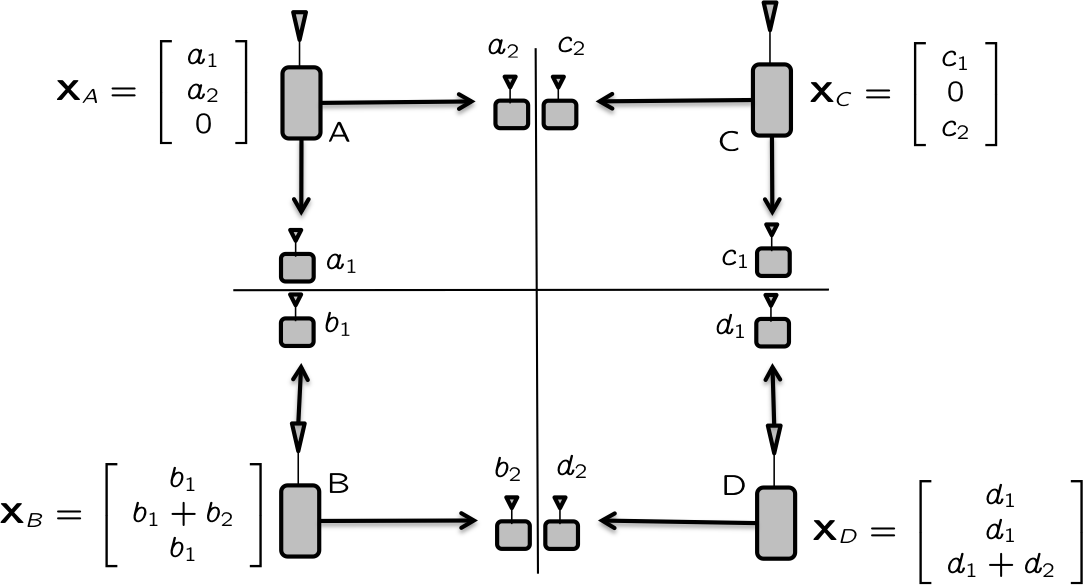}
\caption{\it Transmitted symbols over three time slots. 8/3 DoF are achieved with no CSIT. Orthogonal solutions cannot achieve more than 2 DoF. 8/3  is also the optimal DoF value even with global and perfect CSIT. }
\label{fig:4cellsol}
\end{figure}
\proof
The DoF optimal solution to the motivating example of Fig. \ref{fig:intro} is illustrated in Fig. \ref{fig:4cellsol}. A total of 8/3 DoF are achieved with no CSIT, with each message achieving 1/3 DoF. The transmission takes place over three time slots and the transmitted symbols are shown in Fig. \ref{fig:intro} next to the corresponding transmitters. The following table lists the received signals.

\begin{eqnarray*}
\begin{array}{|l|l|l|l|l|}
\hline\mbox{t}& \mbox{Receivers $a_1, b_1$}&\mbox{Receivers $a_2, c_2$}&\mbox{Receivers $c_1, d_1$}&\mbox{Receivers $b_2, d_2$}\\\hline
1&h_Aa_1+h_Bb_1&h_Aa_1+h_Cc_1&h_Cc_1+h_Dd_1&h_Bb_1+h_Dd_1\\\hline
2&h_Aa_2+h_Bb_1+h_Bb_2&h_Aa_2&h_Dd_1&h_Bb_1+h_Bb_2+h_Dd_1\\\hline
3&h_Bb_1&h_Cc_2&h_Cc_2+h_Dd_1+h_Dd_2&h_Bb_1+h_Dd_1+h_Dd_2\\\hline
\end{array}
\end{eqnarray*}
Note that there is some notational abuse in labeling the receivers by their desired symbols, but we find this notation less cumbersome. For the same reason, AWGN is not explicitly mentioned in the table. In the first column, $t=1,2,3$ represents the three channel uses. The variables $h_A, h_B, h_C, h_D$ represent the channel coefficients from transmitters $A, B, C, D$, respectively, to the receivers indicated at the top of the column. Note that  each receiver at the top of the column has a different channel to the same transmitter. However, since this is non-consequential, we choose the less cumbersome notation where the channels are only identified by the transmitter index. Finally, note that this is clearly not an orthogonal solution.

Since the receivers know the channel coefficients (local CSIR is enough), it is easy to see how each receiver can resolve its desired symbol from its three received signals. Further, note that \emph{both} receivers at the top of the column can resolve each other's messages as well. For instance, receivers $a_1, b_1$ recover symbol $b_1$ from the third time slot and by subtracting it out, obtain the symbol $a_1$ from the first time slot. Interference alignment is evident in the observation that there are only 3 equations available to each receiver, involving 4 unknown (information carrying) symbols. While in general 3 equations cannot be solved for 4 variables, here it is possible to resolve the 2 desired variables at each receiver only because the 2 undesired variables are aligned into one dimension, leaving a two dimensional space free from interference at each receiver, wherein the desired 2 symbols are resolved. Perhaps the most interesting example is the last column where receivers $b_2, d_2$ are able to resolve their desired symbols only because the 2 undesired symbols $b_1, d_1$ are aligned along the same dimension, specifically along the $[1~~~1~~~1]^T$ vector. Equivalently, $b_1, d_1$ always appear together in the same linear combination as $h_Bb_1+h_Dd_1$ and therefore can be treated as a single (aligned) variable, leaving us with 3 equations in 3 variables.

Interestingly, 8/3  is the DoF value, not only with no CSIT, but even with perfect and global CSIT. The proof is identical to the arguments presented in the previous section. Consider any receiver, say $a_1$. Eliminate all messages except $a_1, b_1, b_2$. The single antenna receiver can resolve all three messages (decode desired message $a_1$, remove the signal from $A$, allowing access to ${\bf X}_B$ within finite variance noise distortion, from which $b_1, b_2$ are resolved), so their sum DoF cannot be more than 1. Repeating this argument for all receivers, we have the sum DoF bound of 8/3.\hfill\QED

\subsubsection{Channel Coherence}
A discussion of channel coherence requirements is in order. First, note that a network coherence time of 3 symbols is sufficient for the scheme to work, i.e., if all channels remain constant for 3 symbols, the scheme works fine. Since 3 symbols is not a long duration by practical coherence standards, and we do not distinguish between channels based on coherence times, we are well within the confines of our robust assumptions. Second, note that the coherence is \emph{required} only for the users on the $B-D$ boundary, i.e., receivers $b_2, d_2$. For all other receivers, the scheme works even if channels vary in an arbitrary fashion from each symbol to the next. Also note that receivers $b_1, a_2, c_2, d_1$ require only the channel knowledge to their desired base station transmitters. However, if \emph{all} channels are constant over the 3 symbols, there are additional benefits, as \emph{each} receiver  needs to know the channel coefficient from only its \emph{desired} transmitter. This is because the interfering symbols can simple be subtracted out without learning their channels, e.g., receiver $a_1$ recovers symbol $a_1$ by subtracting the third received signal $h_Bb_1$ from its first received signal $h_Aa_1+h_Bb_1$, and no knowledge of $h_B$ is required.

\begin{figure}[!h]
\centering
\includegraphics[width=5in]{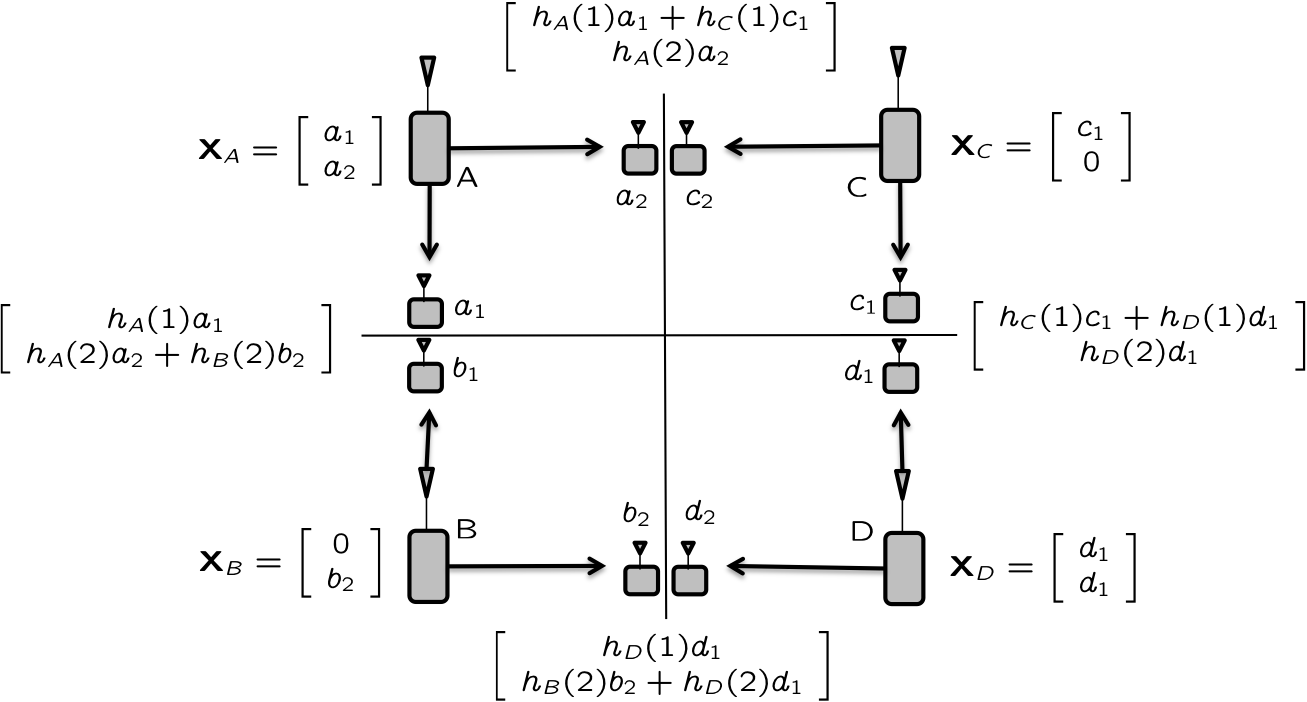}
\caption{\small [Chenwei Wang] {\it Transmitted and received symbols over two time slots. 5/2 DoF are achieved with no CSIT and without any need for channel coherence. Orthogonal solutions cannot achieve more than 2 DoF. }}
\label{fig:4cellsoliid}
\end{figure}

While the assumption of short term network coherence, i.e., that channels do not change over a short duration is quite reasonable, we note that it is not required for all the orthogonal solutions that we found in the previous section. One might wonder if orthogonal solutions would still be optimal if we did not allow any coherence, e.g., if all non-zero channels varied in an i.i.d. fashion. An alternative solution to the same 4 cell example, contributed by Chenwei Wang and shown in Fig. \ref{fig:4cellsoliid}, proves that this is not the case.

Fig. \ref{fig:4cellsoliid} shows the transmitted symbols next to the transmitters and the received signals next to the receivers. $h_A(i)$ denotes the channel coefficient from transmitter $A$ in time slot $i$, and the receiver index is suppressed as before for compact notation. In two channel uses, transmitters $B, C, D$ send one symbol each for a total of 3 symbols $b_2, c_1, d_1$, while transmitter $A$ sends two symbols $a_1, a_2$, to achieve 5/2 DoF. It is easily verified that there is no requirement for the channels to remain constant from one symbol to the next. Since 5/2 DoF is better than the best orthogonal solution, which cannot achieve more than 2 DoF,  it is clear that even under i.i.d. fading with coherence time of unity, orthogonal solutions are not sufficient for DoF optimality.

\subsection{Locally Connected 4 Cell Uplink Network}
\begin{figure}[!h]
\centering
\includegraphics[width=5in]{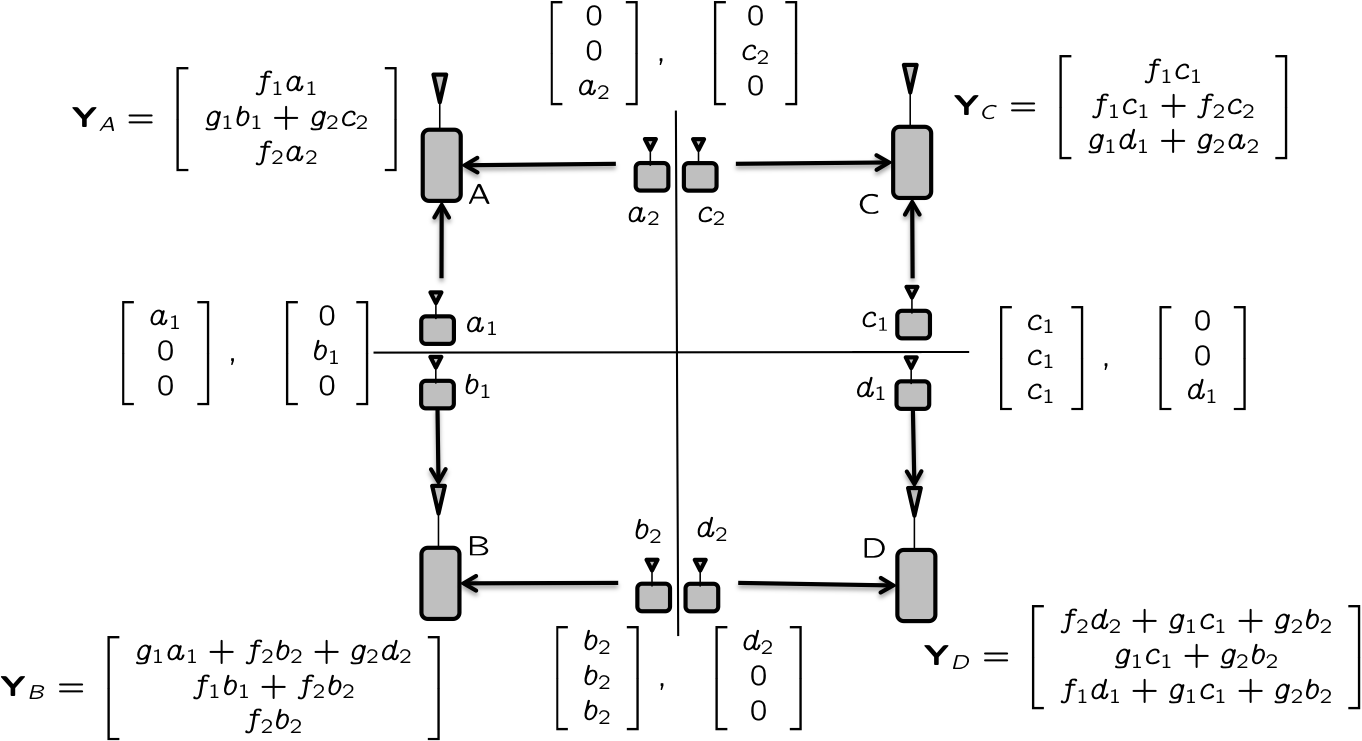}
\caption{{\it (Uplink) Transmitted and received symbols over three time slots. 8/3 DoF are achieved with no CSIT. Orthogonal solutions cannot achieve more than 2 DoF. 8/3 is also the DoF value with perfect and global CSIT.}}
\label{fig:4cellsoldual}
\end{figure}

\begin{figure}[!h]
\centering
\includegraphics[width=5in]{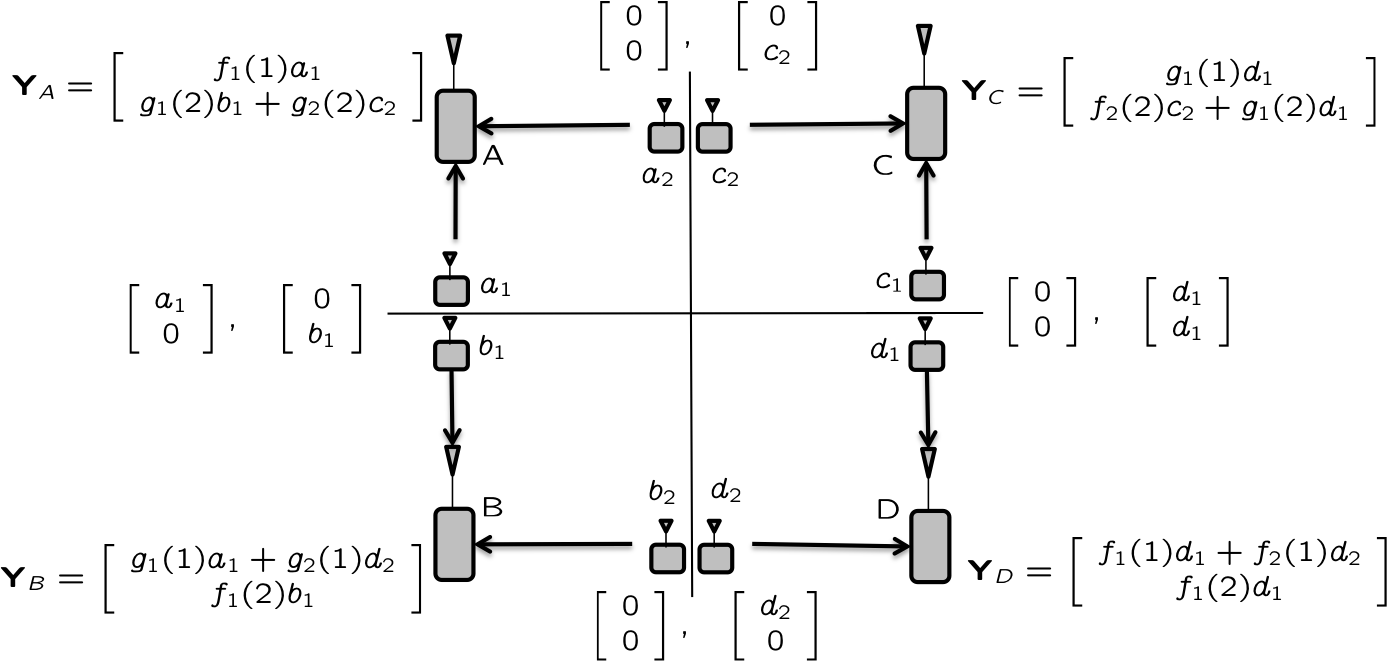}
\caption{{\it Uplink. Transmitted and received symbols over two time slots. 5/2 DoF are achieved with no CSIT and no requirements on channel coherence time. Orthogonal solutions cannot achieve more than 2 DoF. }}
\label{fig:4cellsoliiddual}
\end{figure}

Next we consider the reciprocal setting where the roles of transmitters and receivers are switched for each message, for the 4 cell cluster example, i.e., the uplink setting for our motivating example. For ease of exposition we present all the information in Fig. \ref{fig:4cellsoldual} where the optimal DoF value of 8/3 is achieved with network coherence time of 3 or more symbols, and in Fig. \ref{fig:4cellsoliiddual} where 5/2 DoF are achieved without any constraints on network coherence time.

\begin{theorem}
The 4-cell uplink network, i.e., the reciprocal network of the 4-cell downlink network shown in Fig. \ref{fig:intro}, with sufficient coherence ($\tau_{\max}\geq 3$), has DoF = 8/3, both with or without CSIT, and both with or without spatially i.i.d. fading.
\end{theorem}

Fig. \ref{fig:4cellsoldual} shows the 4 cell uplink solution that achieves 8/3 DoF with no CSIT. The transmitted and received symbols are shown in the figure. Channel variables $f_i$ represent channels to the desired receiver and channel variables $g_i$ represent channels to the undesired receiver. The index $i$ corresponds to the subscript in the transmitter index. So, for example, the channel from transmitter $a_1$ to desired receiver $A$ is $f_1$ and the channel from transmitter $c_2$ to undesired receiver $A$ is $g_2$. The receiver index is suppressed for compact notation, but note that each receiver experiences a different set of channel values, i.e., $f_1$ for receiver $A$ is different from $f_1$ for receiver $C$. The difference in these values is inconsequential, however, since each receiver only processes its own received signals. From the figure, it is easy to verify that receiver $A$ can resolve symbols $a_1, a_2$, receiver $B$ can resolve $b_1, b_2$, $C$ can resolve symbols $c_1, c_2$ and $D$ can resolve symbols $d_1,d_2$. The role of interference alignment is obvious because each receiver sees 3 equations in 4 variables, from which it needs to recover its two desired variables. This can be done only because the 2 undesired variables occupy only a 1 dimensional space, because of interference alignment. A network coherence time of 3 or more symbols is needed. Note also that each receiver needs to know only the channel coefficients for the channels carrying the desired information symbols. On the other hand, if each receiver knows all channels that it can hear, then coherence is only needed for one of the 4 receivers, receiver $D$ in Fig. \ref{fig:4cellsoldual}. To see that 8/3 is also the DoF outer bound the argument is essentially the same as before. Consider any receiver, e.g., receiver $A$ and eliminate all messages except $a_1, a_2, b_1$. The receiver can resolve all three messages ($a_1, a_2$ can be decoded by design and therefore subtracted to reveal $b_1$ within a bounded variance noise distortion), and since the receiver has only 1 antenna, the total DoF of the three messages cannot be more than 1. Repeating the same argument for all receivers, we arrive at the outer bound value of $1/3$  DoF per message, or a total DoF value of 8/3. The argument does not depend on the presence or absence of CSIT, so 8/3  is  the DoF value with or without CSIT.

Finally, Fig. \ref{fig:4cellsoliiddual} shows how 5/2 DoF, which is more than the best possible DoF value with orthogonal solutions, are achievable without CSIT, even if there is no coherence assumption.

\subsection{Locally Connected 4-Cell Network with Spatially i.i.d. Fading: DoF with Cooperation}
\begin{figure}[!h]
\centering
\includegraphics[width=5in]{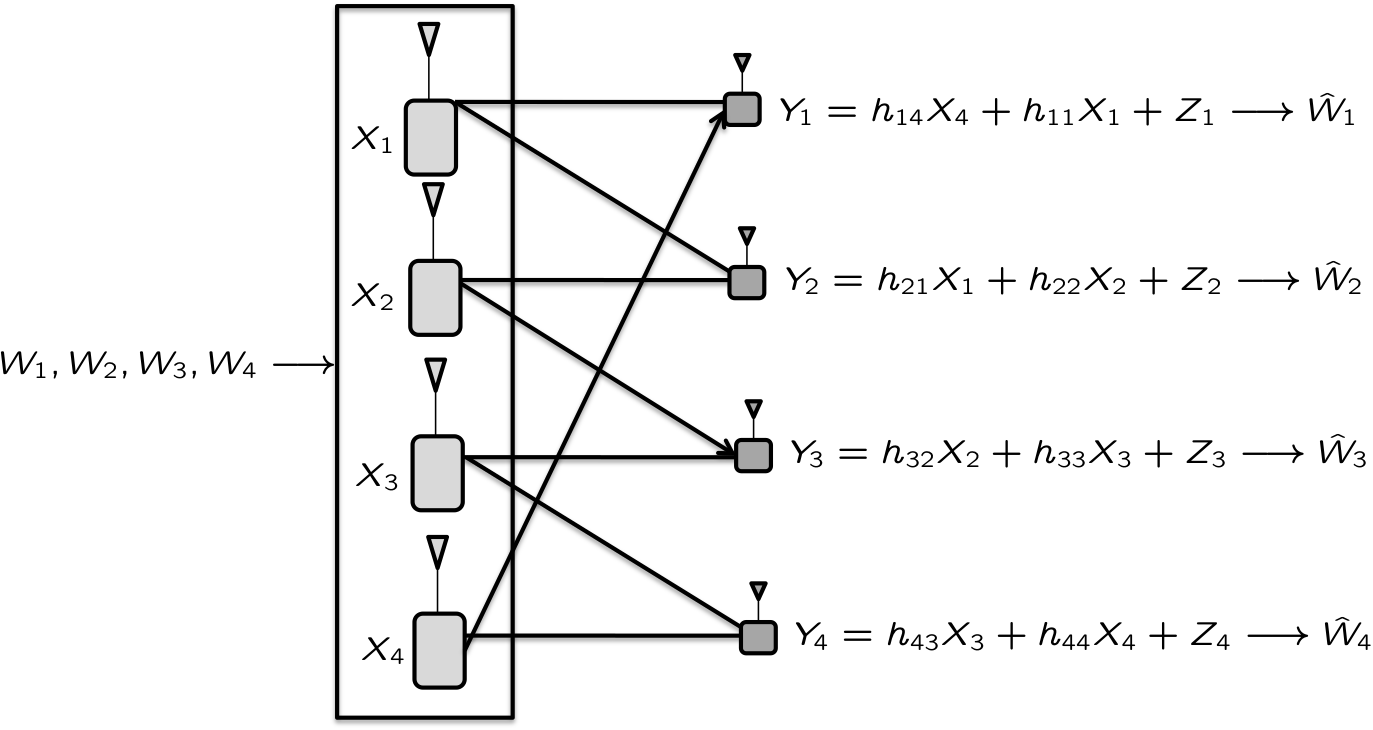}
\caption{{\it Partially connected 4 User MISO BC  with no CSIT, obtained by allowing full cooperation among the 4 transmitters.}}
\label{fig:4cellsolBC}
\end{figure}

Here we explore the impact of transmitter cooperation on the 4-cell downlink network with no CSIT, when all (connected) channels are independent identically distributed, i.e., statistically equivalent. Because of spatially i.i.d. fading, receivers $a_2, c_2$ are equivalent and can be replaced with a single new receiver, say receiver $ca$. Similarly, receivers $a_1,b_1$ are replaced by receiver $ab$, receivers $b_2,d_2$ are replaced with receiver $bd$, and receivers $c_1,d_1$ are replaced with receiver $cd$, producing the partially connected $X$ network of Fig. \ref{fig:intro}. Allowing all transmitters to cooperate and relabeling receivers $ca, ab, bd, cd$ as receivers $1,2,3,4$, respectively, gives us a 4 user MISO BC where the transmitter has 4 antennas and each user has a single antenna, as shown in Fig. \ref{fig:4cellsolBC}. For this BC, we will show a DoF outer bound with no CSIT.

The  result, stated in the following theorem, shows that there is no DoF benefit of transmitter cooperation for this network without CSIT.
\begin{theorem}\label{thm:4cellsolBC}
The locally connected 4-cell network of Fig. \ref{fig:intro} with no CSIT and spatially i.i.d. fading, has DoF $\leq 8/3$ with full transmitter cooperation.
\end{theorem}
{\it Remark:} Note that the outer bound result holds regardless of the network coherence model. On the other hand if network coherence time $\tau_{\max}\geq 3$ then we know that 8/3 DoF are also achievable without cooperation, i.e., with sufficient coherence and spatially i.i.d. fading there is no DoF benefit of transmitter cooperation without CSIT.

\proof To prove the outer bound, we will use the compound setting. Let us add another receiver that wants the message $W_1$ and whose channel is also i.i.d. with that of the original receiver 1. Let us identify the new receiver as $Y'_1$. Similarly, we will add an independent but statistically equivalent copy of receiver 3 and identify it as $Y'_3$. Since the newly added receivers are statistically equivalent to existing receivers and desire the same messages, any achievable scheme for the original network is also an achievable scheme for the compound network. Now that we have a compound setting for receivers 1 and 3, we can even provide full channel knowledge to the transmitter (not a necessary step, but shows the strength of the outer bound), which also cannot reduce capacity. Since none of these operations reduce capacity, the outer bound for the compound channel setting with full channel knowledge is also an outer bound for the original channel. We proceed as follows.

\begin{eqnarray}
nR_1&\leq&I(W_1; Y_1^n)+o(n)\\
&=& h(Y_1^n)-h(Y_1^n|W_{1})+o(n)\\
&\leq& n\log(P)-h(Y_{1}^n|W_{1})+n~o(\log(P))+o(n)\label{eq:a1}
\end{eqnarray}

Similarly, for receiver $Y'_1$ which also wants message $W_1$,
\begin{eqnarray}
nR_1&\leq&n\log(P)-h({Y'}_{1}^n|W_{1})+n~o(\log(P))+o(n)\label{eq:b1}
\end{eqnarray}

Adding (\ref{eq:a1}) and (\ref{eq:b1}), we have
\begin{eqnarray}
2nR_1&\leq&2n\log(P)-h(Y_1^n,{Y'}_{1}^n|W_{1})+n~o(\log(P))+o(n)\label{eq:a1b1}
\end{eqnarray}

Repeating the same steps for receiver $Y_3$ and then $Y'_3$, we have
\begin{eqnarray}
2nR_3&\leq&2n\log(P)-h(Y_3^n,{Y'}_{3}^n|W_{3})+n~o(\log(P))+o(n)\label{eq:c1d1}
\end{eqnarray}
Adding (\ref{eq:a1b1}) and (\ref{eq:c1d1}) we have
\begin{eqnarray}
2n(R_1+R_3)&\leq&4n\log(P)-h(Y_1^n,{Y'}_{1}^n,Y_3^n,{Y'}_{3}^n|W_1,W_{3})+n~o(\log(P))+o(n)\label{eq:c1d1}\\
&=&4n\log(P)-h(X_1^n+\overline{Z}_1^n,X_2^n+\overline{Z}_2^n, X_3^n+\overline{Z}_3^n, X_4^n+\overline{Z}_4^n,|W_1,W_{3})\nonumber\\
&&~~~+n~o(\log(P))+o(n)\\
&=&4n\log(P)-n(R_2+R_4)+n~o(\log(P))+o(n)
\end{eqnarray}
where $\overline{Z}_i\sim \mathcal{N}(0,\mathcal{O}(1))$ and i.i.d. in time. 

\noindent Applying limits $n\rightarrow\infty$ and then $P\rightarrow\infty$, we have
\begin{eqnarray}
2d_1+2d_3+d_2+d_4&\leq& 4
\end{eqnarray}
By symmetry we also have
\begin{eqnarray}
d_1+d_3+2d_2+2d_4&\leq& 4
\end{eqnarray}
which leads us to the sum DoF bound
\begin{eqnarray}
d_1+d_2+d_3+d_4&\leq&\frac{8}{3}
\end{eqnarray}
Thus, cooperation between transmitters does not increase DoF without CSIT, if the network is sufficiently coherent (specifically, $T_{\max}\geq 3$) and connected channels are spatially i.i.d.\hfill\QED

{\it Remark:} Combining the results of Theorem \ref{thm:4cellsol} and Theorem \ref{thm:4cellsolBC}, for the locally connected 4 cell network with spatially i.i.d. fading and sufficient coherence, we have established a curious duality of DoF results:
\begin{enumerate}
\item {\it Without transmitter cooperation, CSIT is not useful (Theorem \ref{thm:4cellsol})}.
\item {\it Without CSIT, transmitter cooperation is not useful (Theorem \ref{thm:4cellsolBC})}.
\end{enumerate}

{\it Remark:} If both CSIT and transmitter cooperation are available, then indeed the DoF are higher. It is easy to verify that in that case the network has 4 DoF almost surely. Similarly, if instead of transmitter cooperation, receiver cooperation is  available, then it is easily verified that the network has 4 DoF almost surely, even without CSIT.

{\it Remark:} Since the dual of the BC of Fig. \ref{fig:4cellsolBC} is itself, we have also established the DoF of the locally connected 4 cell \emph{uplink} network with transmitter cooperation, spatially i.i.d. fading and no CSIT, in the reciprocal setting.

{\it Remark:} The spatially i.i.d. assumption is needed to reduce the 4 cell network into the MISO BC with 4 receivers, i.e., to combine equivalent receivers. However, the DoF outer bound presented above does not require spatially i.i.d. fading. Therefore, for the 4 user partially connected BC shown in Fig. \ref{fig:4cellsolBC}, even without spatially i.i.d. fading, the DoF are bounded above by $\frac{8}{3}$.

\section{Cellular Blind Interference Alignment: a Wireless Index Coding Problem}\label{sec:arbitrary}
In this section we study arbitrarily connected wireless networks, with arbitrary message sets and no CSIT, and relate it to the \emph{index coding} problem, which, incidentally, is the context that produced the first known example of interference alignment \cite{Birk_Kol, Jafar_FnT}. We start with a  description of the {index coding} problem.
\subsection{Index Coding Problem $\mathcal{IC}(\mathcal{W}_r, \overline{\mathcal{W}}_r)$}
Index coding is a network coding problem that is surprisingly easy to describe but can be challenging to solve, and is open in general. It is simply a multiple unicast problem where all links but one have infinite capacity. An index coding problem is defined by the set of receivers, each specified by the message(s) it wants and the messages that are already available to that receiver. Let us define $\mathcal{W}_r$ to be the set of messages desired by receiver $r$, and $\overline{\mathcal{W}}_r$ to be the set of messages already available to receiver $r$. With this notation we refer to an index coding problem as $\mathcal{IC}(\mathcal{W}_r, \overline{\mathcal{W}}_r)$.

\begin{figure}[!h]
\centering
\includegraphics[width=3in]{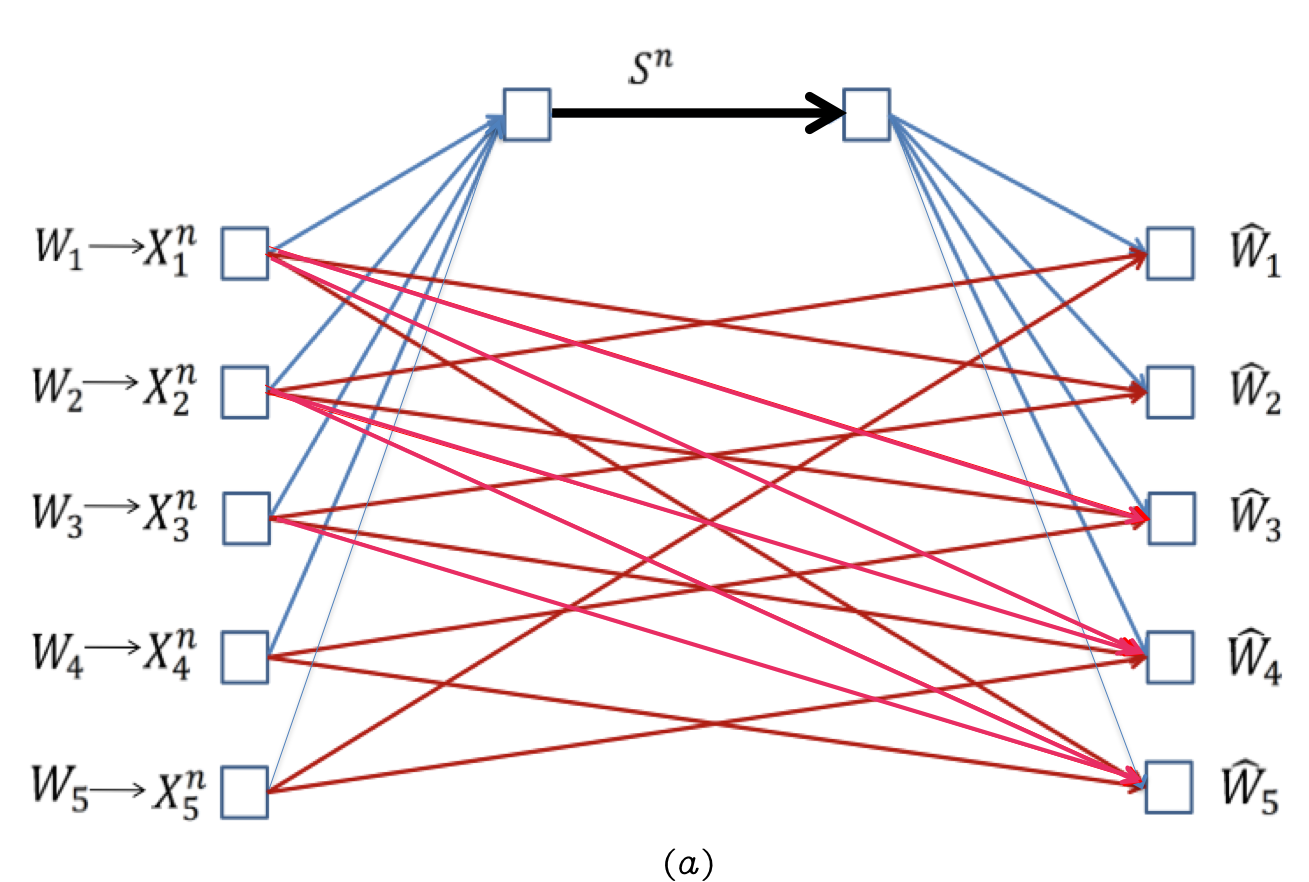}~~~
\includegraphics[width=2.9in]{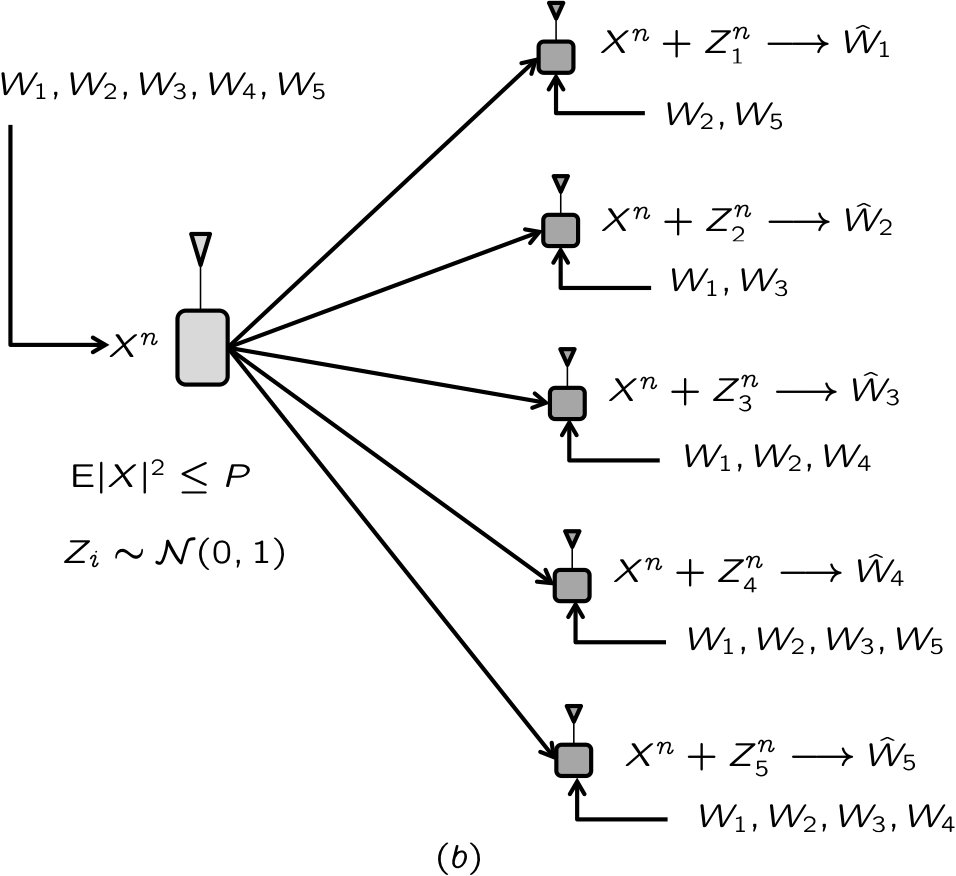}
\caption{{\it (a) Index coding problem. (b) Corresponding DoF problem: Gaussian index coding}}
\label{fig:index}
\end{figure}

An example is shown in Fig. \ref{fig:index} where we have 5 messages $W_1,\cdots, W_5$. Only the link shown in black at the top of the figure has finite capacity, which maybe assumed to be unity without loss of generality. All other links have infinite capacity. This means that receiver 1, who wants $W_1$, already has knowledge of $W_2, W_5$. Receiver 2, who wants $W_2$, has knowledge of messages $W_1, W_3$. Receiver $3$ wants message $W_3$ and has knowledge of $W_1, W_2, W_4$. Receiver $4$ wants $W_4$ and knows $W_1, W_2, W_3, W_5$. Receiver $5$ wants $W_5$ and knows $W_1, W_2, W_3, W_4$. Equivalently
\begin{eqnarray*}
\mathcal{W}_1=\{W_1\}, &&\overline{\mathcal{W}}_1=\{W_2,W_5\}\\
\mathcal{W}_2=\{W_2\}, &&\overline{\mathcal{W}}_2=\{W_1,W_3\}\\
\mathcal{W}_3=\{W_3\}, &&\overline{\mathcal{W}}_3=\{W_1,W_2,W_4\}\\
\mathcal{W}_4=\{W_4\}, &&\overline{\mathcal{W}}_4=\{W_1,W_2,W_3,W_5\}\\
\mathcal{W}_5=\{W_5\}, &&\overline{\mathcal{W}}_5=\{W_1,W_2,W_3,W_4\}
\end{eqnarray*}

Shown to the right in Fig. \ref{fig:index} is the corresponding DoF problem for a Gaussian broadcast channel with cognitive receivers, which we label as $\mathcal{GIC}(\mathcal{W}_r, \overline{\mathcal{W}}_r)$. In spite of the difference of finite fields versus real/complex fields,  the two are  equivalent (i.e., have the same capacity) if the capacity of the bottleneck edge in the index coding problem is chosen to be the same as the capacity of the Gaussian link in the Gaussian index coding problem. The equivalence is readily seen as a direct consequence of the much more general ``network equivalence" result of Koetter et al. \cite{Koetter_Effros_Medard}. Interestingly, in the network coding formulation over finite fields, it is known that linear coding is not always sufficient \cite{Blasiak_Kleinberg_Lubetzky_2011}. While it seems that the equivalence of the two networks might imply a similar insufficiency result for  the Gaussian index coding problem, to the best of our knowledge, such a result has not been formally established. Note that by linear coding in the Gaussian index coding problem, we mean the DoF achievable through linear beamforming across multiple channel uses. In the Gaussian index coding problem it is  possible to perform network coding over finite fields  at the bit level, i.e., a network coding of messages, entirely before the channel coding for transmission over the Gaussian channel. The channel coding in that case only helps to convert the Gaussian channel into a noise-free data pipe, making the Gaussian index coding problem  identical to the index coding problem. For such bit level network coding it is already known that linear schemes are not sufficient. However, the linearity that is of interest to us, is where each message is independently linearly pre-coded into beamforming vectors (over complex numbers) at the physical layer within the vector space created by multiple channel uses and the separability of interfering messages is only guaranteed by linear independence of these vectors. The reason for our interest in  linear precoding at the physical layer will soon be evident. It is the Gaussian index coding formulation that is of immediate interest to us.

\subsection{Cellular blind interference alignment as a wireless index coding problem}
Let us define a mapping from a cellular blind interference alignment problem to a (Gaussian) index coding problem.

{\bf Mapping from CB to GIC:} To each cellular blind interference alignment problem $\mathcal{CB}(\mathcal{C},\mathcal{W}_t, \mathcal{W}_r)$ let us associate the Gaussian index coding problem $\mathcal{GIC}(\mathcal{W}_r, \overline{\mathcal{W}}_r)$, such that the sets of desired messages $\mathcal{W}_r$ are the same in both problems and 
\begin{eqnarray}
\forall r\in\mathcal{R},~~~\overline{\mathcal{W}}_r&=&\bigcup_{\small\begin{array}{c}t\in\mathcal{T}\\c_{rt}=0\end{array}}\mathcal{W}_t\label{eq:Wrbar}
\end{eqnarray}
In other words, the set of messages already available to a receiver in the (Gaussian) index coding problem is simply the set of all the messages in the CB problem that originate at transmitters which are not connected to this receiver.

{\bf Mapping from GIC to CB:} To each  Gaussian index coding problem $\mathcal{GIC}(\mathcal{W}_r, \overline{\mathcal{W}}_r)$, let us associate the cellular blind interference alignment problem $\mathcal{CB}(\mathcal{C},\mathcal{W}_t, \mathcal{W}_r)$ such that the sets of desired messages $\mathcal{W}_r$ are the same in both problems and 
\begin{eqnarray}
\mathcal{W}_t&=&\{W_t\}\\
c_{rt}&=&\left\{\begin{array}{lll}
0&& \mbox{ if } W_t\in\overline{\mathcal{W}}_r\\
1&& \mbox{ otherwise}.\end{array}\right.
\end{eqnarray}
In other words, there is a separate transmitter for each message, and a transmitter-receiver pair are not connected in the CB problem if and only if the corresponding message was available a-priori to the receiver in the GIC problem.

The following theorems formalize the general relationship between  cellular blind interference alignment problems and index coding problems.
\begin{theorem}\label{thm:cbic}
The DoF of each cellular blind interference alignment problem under the assumption of spatially i.i.d. fading, are bounded above by the DoF of the corresponding Gaussian index coding problem.
\begin{eqnarray}
DoF\left(\mathcal{CB}(\mathcal{C},\mathcal{W}_t, \mathcal{W}_r)\right)&\leq&DoF\left(\mathcal{GIC}(\mathcal{W}_r, \overline{\mathcal{W}}_r)\right)
\end{eqnarray}
where  $\overline{\mathcal{W}}_r$ is defined in (\ref{eq:Wrbar}).
\end{theorem}
\proof Starting from the CB problem, we will go through a series of steps, each of which cannot reduce DoF, to arrive at the corresponding GIC problem. 
\begin{enumerate}
\item In the CB problem, let a Genie provide the messages $\overline{\mathcal{W}}_r$ to receiver $r$, $\forall r\in\mathcal{R}$.
\item Allow full connectivity, $c_{rt}=1$, ~$\forall t\in\mathcal{T}, r\in\mathcal{R}$. This cannot hurt because of the previous step which provided all the messages from the ''formerly disconnected" transmitters to each receiver. Knowing all the messages from a transmitter allows the receiver to construct the transmitted codewords. Since the receiver has full CSIR, it can simply subtract the new interferers.
\item Allow full cooperation between all transmitters. This gives us a MISO BC with cognitive receivers, i.e., where some receivers have prior  knowledge of some messages.
\item Since the capacity of the BC depends only on the marginals and the channels are spatially i.i.d., without loss of generality let us make the channel vector for one of the users (say, the first user), also the channel vector of all the users.
\item Let us provide full CSIT to the transmitter.
\item Since the transmitter has full CSIT, and the overall channel matrix (from the transmitter to all receivers) has only unit rank, the transmitter can do a change of basis operation to discard all redundant dimensions, leaving the transmitter with a single antenna.
\item Without loss of generality for DoF metric, both the channel gain and noise variance at each receiver can be normalized to unity, e.g., by scaling at the transmitter, or the receivers. This is only a cosmetic step, since the DoF value is unaffected by these normalizations.
\end{enumerate}
The problem thus obtained is the corresponding GIC problem\hfill\QED

\begin{theorem}\label{thm:iccb}
The maximum DoF value achievable through linear beamforming in the Gaussian index coding problem is equal to the maximum DoF value achievable through linear beamforming in the corresponding cellular blind interference alignment problem under the assumption of sufficient coherence (i.e., assuming network coherence interval, $\tau_{\max}$, is large enough).
\end{theorem}
\proof A linear beamforming solution involves simultaneous transmission of information symbols from each message along different (beamforming) vectors, in the vector space created by multiple channel uses and complex dimensions. The simultaneous transmissions are linearly superimposed on each other. For a linear beamforming solution to be feasible, each receiver must be able to resolve its desired symbols. Specifically, what is required is that the vectors carrying the desired symbols at each receiver, be linearly independent of each other and of the vectors carrying the undesired symbols (interference). 

Given a linear beamforming solution for the GIC problem, we can use the same beamforming vectors for the CB problem as well. Since the transmitters are distributed in the CB problem, the beamforming vectors are assigned to the transmitters where the corresponding messages originate. Simultaneous transmission by all transmitters creates the same superposition of signal dimensions as in the GIC problem. While each transmission gets scaled by the corresponding (non-zero) channel coefficient in the CB problem, note that because we assume sufficient coherence, the channel coefficient remains constant throughout. Finally, since scaling the beamforming vectors by  non-zero constants does not affect  the linear independence properties of the set of vectors, a feasible solution for the GIC problem is automatically a feasible solution for the CB problem as well.\hfill\QED

\begin{theorem}\label{thm:linear}
Whenever linear beamforming is DoF optimal for the Gaussian index coding problem it is also optimal for the corresponding cellular blind interference alignment problem under the assumption of sufficient coherence and spatially i.i.d. fading.
\end{theorem}

\proof The result is a direct consequence of Theorem \ref{thm:cbic} and Theorem \ref{thm:iccb}. \hfill\QED

\subsection{Understanding the relationship between CB and GIC}
The relationship between CB (with sufficient coherence and spatially i.i.d. fading) and GIC, formalized by theorems \ref{thm:cbic}, \ref{thm:iccb}, \ref{thm:linear}, is evidently quite strong. What these theorems imply is that when linear solutions are optimal, the two DoF problems are equivalent. Note that the insufficiency of linear solutions has been an open question even for index coding over finite fields, until recently a gap between linear and non-linear solutions was established by Blasiak et al. \cite{Blasiak_Kleinberg_Lubetzky_2011}. While one might expect this result to carry over to the Gaussian index coding problem as well, to the best of our understanding, it is  not yet established that linear beamforming solutions are insufficient for achieving the DoF of the Gaussian index coding problem. In any case, even if we set aside the question of  sufficiency of linear beamforming schemes, the complexity of settling this issue does say something about the power of linear beamforming schemes. In other words, even if linear beamforming schemes are not always DoF optimal, it is amply evident that they are quite often DoF optimal. The equivalence between CB and GIC can also be interpreted in the same sense.

There is also an important distinction between CB and GIC settings. As mentioned earlier, the GIC setting can take advantage of index coding solutions over finite fields by simply performing them at the bit level before the physical layer coding,  i.e., alignment of message bits rather than alignment of physical signals. Physical layer channel coding would then be used simply to clean up the noisy channel, i.e., to create an effective noise-free data pipe over which aligned message bits can be sent to all receivers. This does not, however, appear to be possible for the CB setting, because of two reasons. First, the messages coming from the same transmitter cannot be aligned. This is because in the CB setting, no receiver has prior knowledge of a strict subset of the messages originating at any particular transmitter. Either the receiver knows all the messages (equivalently, it is not connected to the transmitter), or it knows none of them (equivalently, it is connected) a-priori. Second, \emph{message} alignments from multiple transmitters is not possible because, the transmitters are distributed, carry disjoint message sets, and therefore cannot jointly pre-process the messages, and also because the superposition of signals takes place through unknown scaling factors (channel coefficients) which would appear to rule out the possibility of lattice alignment. Because of these differences, we conclude that unlike the GIC problem which may be essentially the same as an index coding problem and can certainly access index coding solutions over finite fields, \emph{the CB problem is truly a ``wireless" index coding problem}, and the distinction, as explained above, is quite significant. We will also see further evidence of this distinction in the section on interference diversity.

\subsection{Solving the Wireless Index Coding Problem}
Having recognized the cellular blind interference alignment problem as a wireless index coding problem, the natural next question is how to use this new insight to find solutions to the cellular blind interference alignment problem. Indeed, the index coding problem is itself an open problem. In fact, it is known that in its generality the index coding problem is rich enough to capture the entirety of the multiple unicast problem in network coding \cite{Rouayheb_Sprintson_Georghiades}, which points to the difficulty of obtaining  systematic solutions for  blind cellular interference alignment with arbitrary connectivities and arbitrary message sets. However, approximations, heuristic approaches and even optimal solutions for special cases, e.g., symmetric network settings \cite{Blasiak_Kleinberg_Lubetzky_2010, Maleki_Cadambe_Jafar} or small networks, are often readily available, tractable, even quite simple as they are mostly just linear beamforming solutions. Consider for example, our aligned frequency-reuse solutions, which can also be posed as index coding problems, and turn out to be quite simple.

\begin{figure}[!h]
\centering
\includegraphics[width=2.9in]{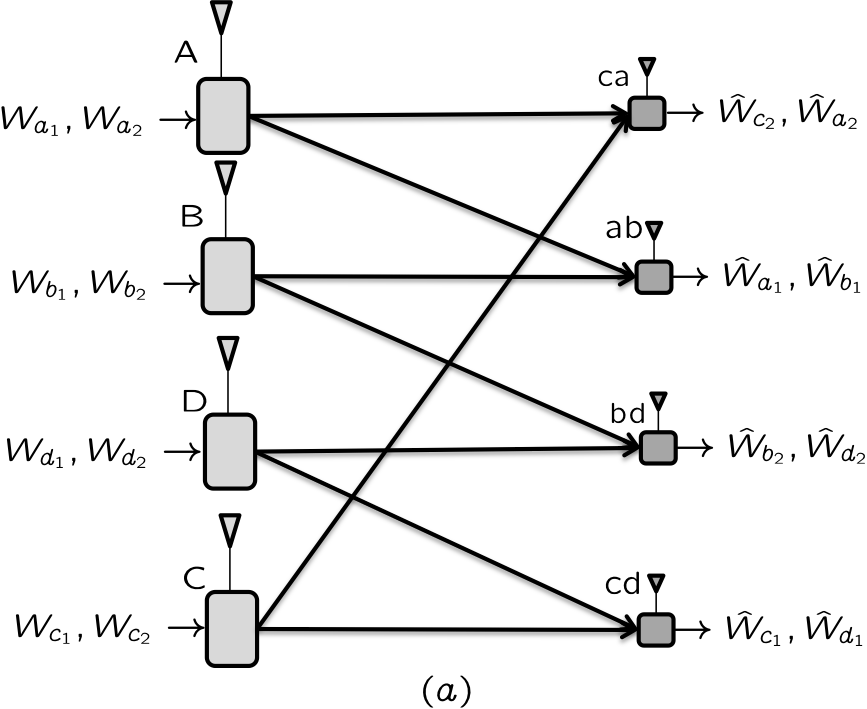}~~~~\includegraphics[width=2.9in]{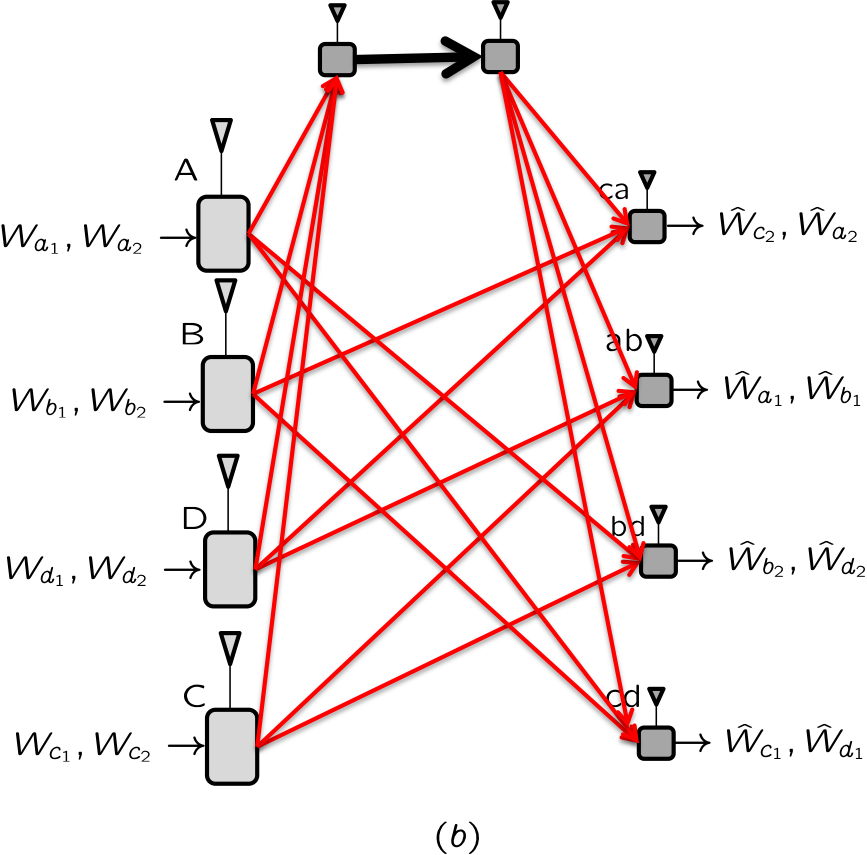}\\\vspace{0.5cm}\includegraphics[width=3.5in]{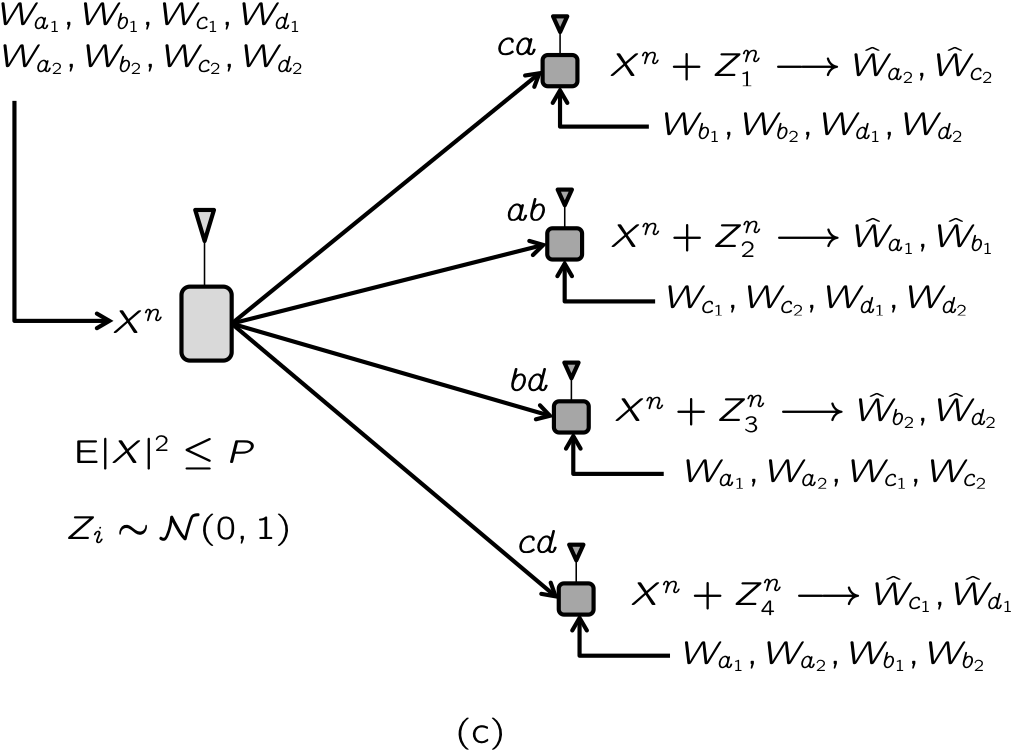}~~~\includegraphics[width=2.3in]{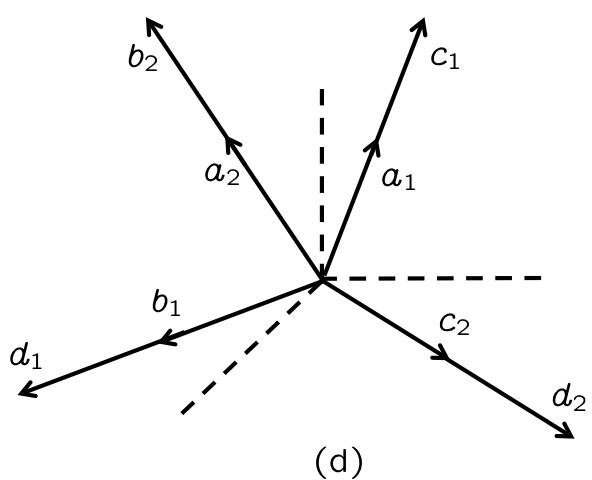}
\caption{\it (a) Locally connected 4 Cell network problem as a partially connected X network, (b) Corresponding index coding problem, (c) Corresponding Gaussian index coding problem, (d) Interference alignment solution: 4 beamforming vectors are used in a 3 dimensional space, each carrying two symbols, for a total of 8/3 DoF.}
\label{fig:4cellindex}
\end{figure}
To illustrate some of the intuition that guides the solutions to index coding problems, we re-visit our locally connected 4-cell network problem. Under the spatially i.i.d. fading assumption, the CB problem is represented by the X network shown in Fig. \ref{fig:4cellindex}(a). To convert this into the corresponding index coding problem shown in Fig. \ref{fig:4cellindex}(b), we first take the complement of the CB network graph, i.e., we invert the connectivity (any transmitter and receiver not connected in the CB problem are connected in the IC problem and any transmitter and receiver that are connected in the CB problem are not connected in the IC problem), and then add the two nodes with the bottleneck link at the top, connecting every transmitter to the source of the bottleneck and every destination to the destination of the bottleneck. The GIC problem is obtained by the mapping described previously and shown in Fig. \ref{fig:4cellindex}(c). Let us focus on this figure. To find the solution we list the desired alignments at each receiver. Receiver $ca$ sees interference from messages $a_1, c_1$ because these are the messages neither desired by receiver $ca$, nor available to the receiver $ca$. Alignment of $a_1,c_1$ would be desirable for receiver $ca$, because it would free more interference free dimensions for this receiver. Similarly, receiver $ab$ would like $a_2,b_2$ to align, receiver $bd$ would like $b_1, d_1$ to align, and receiver $cd$ would like $c_2, d_2$ to align. Thus we have four alignment groups $(a_1, c_1), (a_2, b_2), (b_1,d_1), (c_2,d_2)$. Since each receiver desires two messages, and the aligned interference must occupy at least as much space as one message, the total signal space must be at least three times the size of the space assigned to each message. Therefore, a natural solution is to operate over a 3 dimensional space, and assign one dimension to each message. Further, to achieve interference alignment the dimensions assigned to $(a_1, c_1)$ should be the same, as should be the dimensions assigned to $(a_2, b_2)$, $(b_1, d_1)$, $(c_2, d_2)$. Thus, the solution is simply to choose 4 pairwise linearly independent vectors in a 3 dimensional space and assign one to each of the 4 groups, as shown in Fig. \ref{fig:4cellindex}(d).  The solution shown in Fig. \ref{fig:4cellsol} assigns the vector $[1, 0, 0]^T$ to $(a_1, c_1)$, the vector $[0, 1, 0]^T$ to $(a_2, b_2)$, the vector $[0,0,1]^T$ to $(c_2,d_2)$, and the vector $[1,1,1]$ to $(b_1,d_1)$.  Note that any 4 pairwise linearly independent vectors would have sufficed, but the columns of the identity matrix are preferable, because they do not require coherence of channels. As mentioned previously, the solution presented in Fig. \ref{fig:4cellsol} requires coherence only for the receiver $bd$ where the alignment vector $[1,1,1]^T$ is used.  By the same token, note that the solution presented in Fig. \ref{fig:4cellsoliid} which requires no coherence, uses only the columns of the identity matrix for alignment. The vector $[1,1]^T$, used for $d_1$ does not align with any other symbol. In general, faced with coherence interval constraints of $\tau_{\max}$ alignment can evidently only be achieved over vectors whose support (i.e., the locations of non-zero elements) is limited to $\tau_{max}$ adjacent elements. Alignment solutions with these additional support constraints are a promising direction for further study of robust cellular blind interference alignment principles.

To conclude this section, we point out that optimal index coding solutions are already known for many settings, and are linear beamforming solutions with few exceptions. These solutions directly carry over to the corresponding CB problem as DoF optimal solutions. For example, the necessary and sufficient condition for each message in an index coding problem to simultaneously achieve rate 0.5, or equivalently for every message in a CB problem to simultaneously achieve 0.5 DoF is identified in \cite{Maleki_Cadambe_Jafar, Blasiak_Kleinberg_Lubetzky_2010}, and turns out to be a simple test of consistency of required alignments. For instance, in Fig. \ref{fig:index} a rate of 0.5 per message is not achievable because, receiver 1 would like $W_3, W_4$ to align and receiver 2 would like $W_4,W_5$ to align, which means all 3 messages $W_3, W_4, W_5$ should align. However, receiver 3 wants message $W_3$ and does not have prior knowledge of message $W_5$, so if $W_3, W_5$ were aligned, receiver 3 would not be able to separate them, thus making the alignment requirements inconsistent and therefore 0.5 DoF per message infeasible. 

Similarly, optimal index coding solutions to symmetric settings are found in \cite{Maleki_Cadambe_Jafar} and turn out to be linear, thereby establishing the optimal DoF for the corresponding CB problem with i.i.d. spatial fading and sufficient coherence. To be concrete, let us present the DoF optimal solution for a class of CB problems that follows directly from the corresponding solution in \cite{Maleki_Cadambe_Jafar}.  Let us define the problem, parameterized by variables $D, U, K$ as follows.

\begin{eqnarray}
\mathcal{T}&=&\{t_1,t_2,\cdots,t_K\}\\
\mathcal{R}&=&\{r_1,r_2,\cdots,r_K\}\\
\mathcal{W}&=&\{W_1,W_2,\cdots,W_K\}\\
\mathcal{W}_{t_k}&=&\{W_k\}\\
\mathcal{W}_{r_k}&=&\{W_k\}\\
\mathcal{C}&=&\{(r,t) \mbox{ such that } t\notin \{r\pm 1, r\pm 2, \cdots, r\pm U, r+U+1,\cdots, r+D\}\}
\end{eqnarray}
where indices are interpreted in a `mod $K$' fashion so that, e.g., $(r,t)=(0, 0)$ is the same as $(r,t)=(K, K)$.  In other words, we have $K$ transmitters, $K$ receivers, $K$ messages, the $k^{th}$ message originates at the $k^{th}$ transmitter and is intended for the $k^{th}$ receiver,  each receiver is not connected to the $U$ transmitters before it, and the $D$ transmitters after it, and is connected to all others. For this setting, the information theoretic DoF value, as well as the linear achievability proof, follows directly from the result of \cite{Maleki_Cadambe_Jafar} as  $\frac{U+1}{K-D+U}$ per message.   
This is achieved by sending $U+1$ symbols for each message over $K-D+U$ dimensions. 

\begin{figure}[!h]
\centering
\includegraphics[width=4.5in]{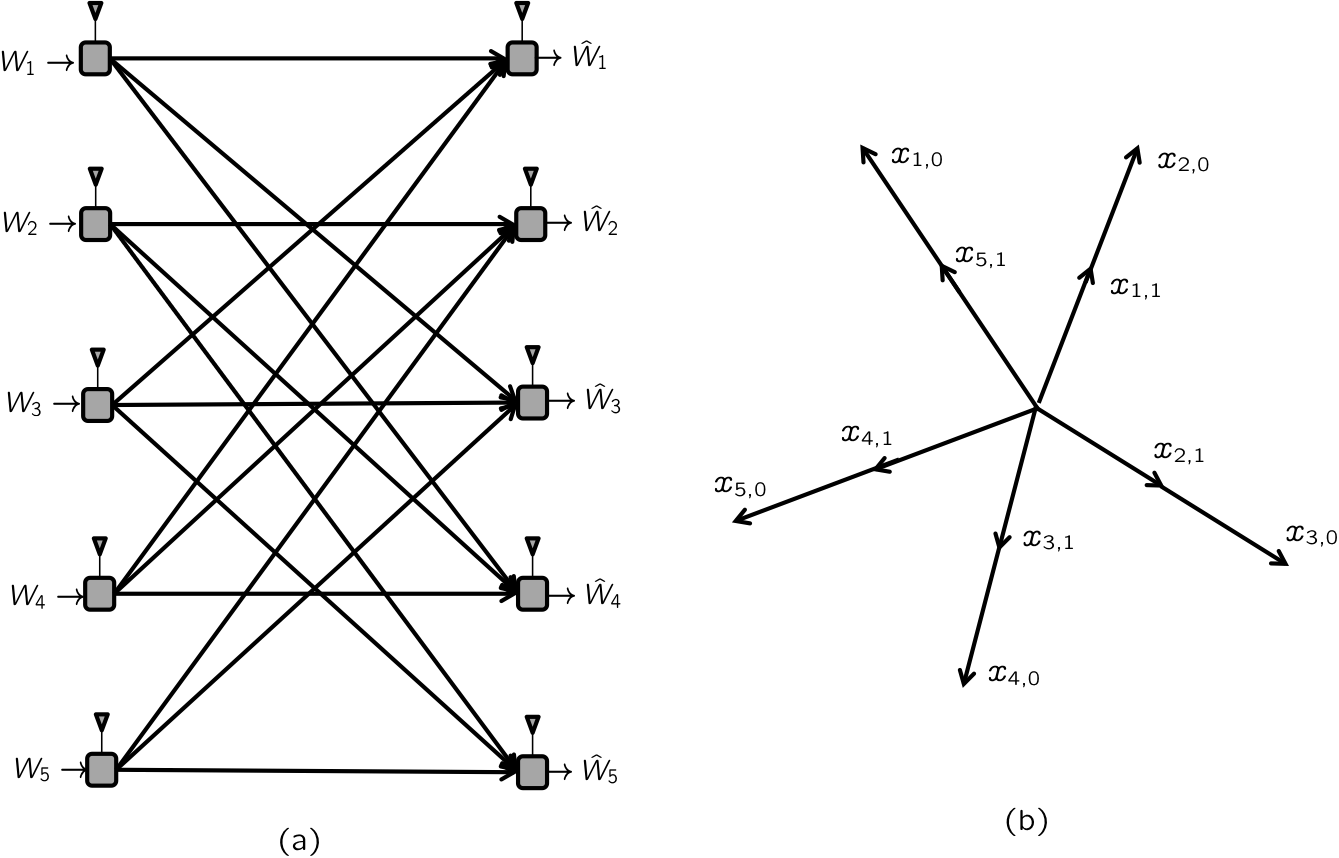}
\caption{\it  Symmetric index coding problem setting of \cite{Maleki_Cadambe_Jafar} with $U=1, D=1, K=5$, applied to the CB problem. (a) The problem, (b) The Solution. In this case no coherence is required.}
\label{fig:index5}
\end{figure}

Fig. \ref{fig:index5}(a) illustrates the problem setting for $U=1, D=1, K=5$. Fig. \ref{fig:index5}(b) shows the interference alignment solution, where two symbols $x_{k,0}, x_{k,1}$ are sent for each message $W_k$, over a 5 dimensional space. This is accomplished by choosing 5 linearly independent alignment vectors over the 5 dimensional space. Since we can choose the five columns of the identity matrix as the 5 linearly independent vectors, each vector can simply be interpreted as a distinct time slot. With this interpretation, in time $k$, source $k$ sends the symbol $x_{k,0}$ and source $k-1$ sends the symbol $x_{k-1, 1}$, so that because of the connectivity, each receiver only hears its desired symbols and no interference. Interestingly, in this case no coherence is required. Note that in general the linear achievability scheme of \cite{Maleki_Cadambe_Jafar} requires alignments along $K$ linearly independent vectors in a $K-D+U$ dimensional space, so when $D>U$, sufficient coherence is required.

\section{Interference Diversity}
Finally, we illustrate one more important element of cellular blind interference alignment: interference diversity. Interference diversity  refers to the observation that  each receiver experiences a different set of interferers, and therefore depending on the actions of its own set of interferers, the interference-free signal space at each receiver fluctuates differently from other receivers. The knowledge of these pre-determined fluctuations, without requiring CSIT, creates  opportunities for blind interference alignment schemes  in the manner of  \cite{Jafar_corr, Wang_Gou_Jafar, Wang_Gou_Jafar_MIMO}, but without the caveats of these previous works. 

\begin{figure}[!h]
\centering
\includegraphics[width=4.5in]{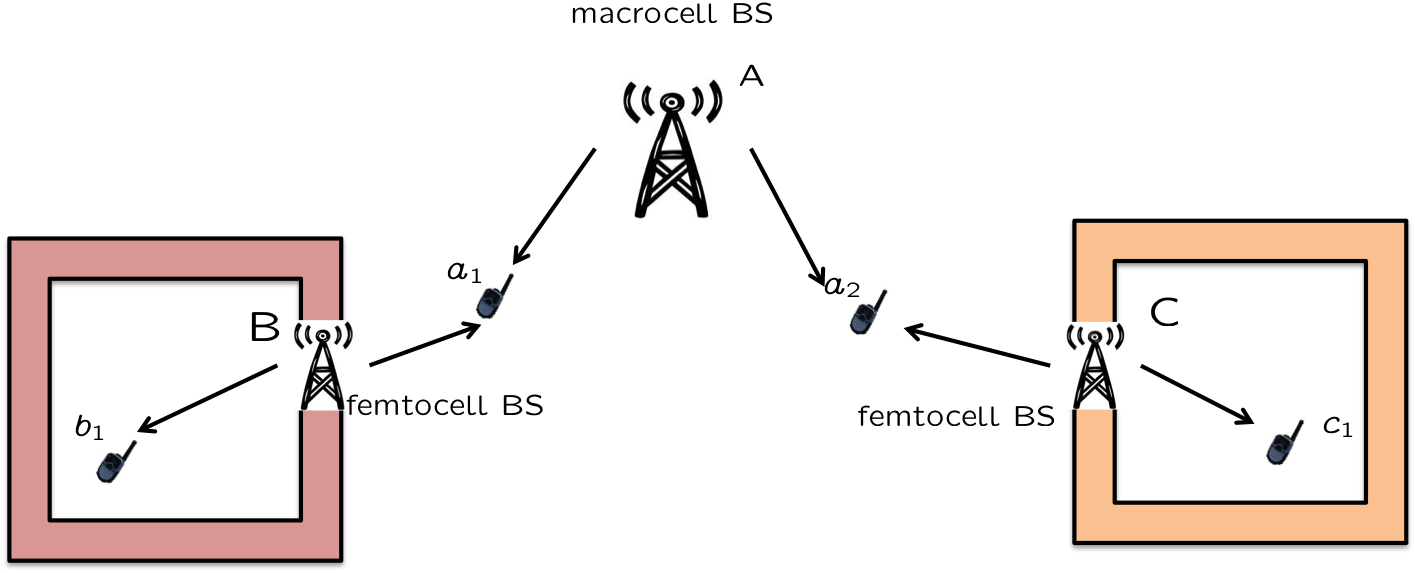}
\caption{\small\it Interference to macrocell users from in-band femtocell deployments}
\label{fig:macrofemto}
\end{figure}
We illustrate this concept through an example of a heterogeneous cellular downlink, shown in Fig. \ref{fig:macrofemto}. Such a situation would be typical for customer-deployed cells such as femtocells. For example, consider customers in areas around $b_1$ and $c_1$, who are located inside macrocell dead spots and set up their own femtocell base stations $B, C$ for wireless access, but then these femtocell base stations also interfere with neighboring macrocell users $a_1, a_2$, respectively. 

\begin{figure}[!h]
\centering
\includegraphics[width=4.5in]{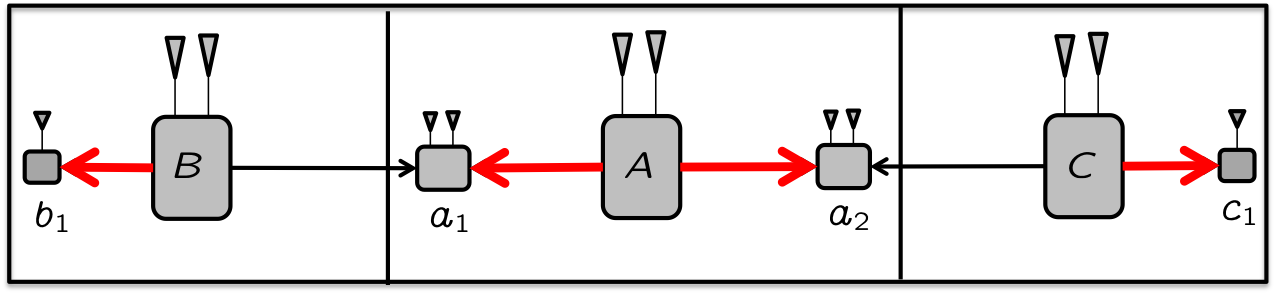}
\caption{\small\it Cellular Blind Interference Alignment setting exploiting Interference Diversity. Cell $A$ can achieve 4/3 DoF while Cells B and C achieve 1 DoF each, with no CSIT, spatially i.i.d. fading, identical physical coherence patterns for all users, no reconfigurable antennas,  and with sufficient coherence ($\tau_{\max}\geq 3$)}
\label{fig:3cellMIMO}
\end{figure}

The resulting CBIA problem is equivalently shown in Fig. \ref{fig:3cellMIMO} with asymmetric antenna configurations.   Note that while all transmitters (base stations) are equipped with two antennas, the receivers in cells $B, C$ are equipped with only single receive antennas.  We  assume spatially i.i.d. fading and sufficient coherence for this example. Since the goal of this example is to highlight interference diversity, we will focus on cell $A$, and the two receivers $a_1, a_2$, who experience different interferers in transmitters $B, C$, respectively. {\it Suppose cells B, C achieve 1 DoF each. The question is to find out how many DoF  cell $A$ can achieve simultaneously.} A naive argument  might be as follows. Since cells $B, C$ achieve 1 DoF each and have only single antenna receivers, the interference from base stations $B, C$ must be consume one DoF  at each of receivers $a_1, a_2$, respectively. Essentially, this would mean that receivers $a_1, a_2$ would sacrifice one antenna to cancel the interference. That leaves cell A with a two-antenna transmitter, two single-antenna receivers, no CSIT, and spatially i.i.d. fading, which would suggest a collapse of DoF for cell $A$, i.e., cell $A$ should not be able to achieve any more than $1$ DoF. 

The naive reasoning is incorrect because it ignores the diversity of interference seen by the two receivers $a_1, a_2$. It turns out that cell $A$ can achieve $4/3$ DoF with no CSIT, spatially i.i.d. fading, no reconfigurable antennas, identical physical coherence models for all channels, and  sufficient coherence $(\tau_{\max}\geq 3)$. The achievable scheme is simply the blind alignment scheme of \cite{Jafar_corr}, albeit without the restriction that nature must produce suitable distinct coherence patterns, and without the restriction to reconfigurable antennas as in \cite{Wang_Gou_Jafar}. Instead, the key here is interference diversity. We operate over three time slots. Transmitters $B, C$ send a new symbol in each time slot. However, transmitter $B$ uses his first transmit antenna for the first 2 time slots and then his second transmit antenna for the third time slot, whereas transmitter $C$ uses his first transmit antenna for the first time slot and his second transmit antenna for the last 2 time slots. In cell $A$, receivers $a_1, a_2$ simply discard the dimension along which interference is received. The net effect is that receiver $a_1$ becomes  a single antenna receiver, with a channel that remains the same for the first 2 time slots and then changes during the third time slot, whereas receiver $a_2$ becomes a single antenna receiver with a channel that changes after the first time slot and then stays constant across the last two time slots. In other words, we have created the staggered coherence blocks required for blind interference alignment \cite{Jafar_corr}, and thus, 4/3 DoF are easily achieved by cell $A$, but without requiring reconfigurable antennas, or relying on nature to create suitable statistical distinctions between the users. However, somewhat curiously, a tight DoF outer bound for this example has so far been elusive and the optimality of $4/3$ DoF for cell $A$ remains open. Perhaps even more curiously, it can be shown that cell A can achieve 2 DoF while cells B, C achieve 1 DoF each, if the channel uncertainty is restricted to choices from finite cardinality sets of generic values for each channel coefficient, i.e., in the finite state compound setting.

The interference diversity example suggests that the cellular blind interference alignment problem, in spite of its similarity to the index coding problem in homogeneous settings, can present a distinct set of challenges in heterogeneous settings with arbitrary antenna configurations on top of arbitrary connectivity and arbitrary message sets.  We further highlight this distinction in the next section.

\subsection{Index Coding versus Wireless Index Coding}
In Section \ref{sec:arbitrary} we established the cellular blind interference alignment problem as a wireless index coding problem, in that it is similar to the index coding problem, but not quite the same due to its wireless character. Here we further emphasize the latter distinction between index coding and wireless index coding (cellular blind interference alignment), through the example of Fig. \ref{fig:3cellMIMO}. 

\begin{figure}[!h]
\centering
\includegraphics[width=2.5in]{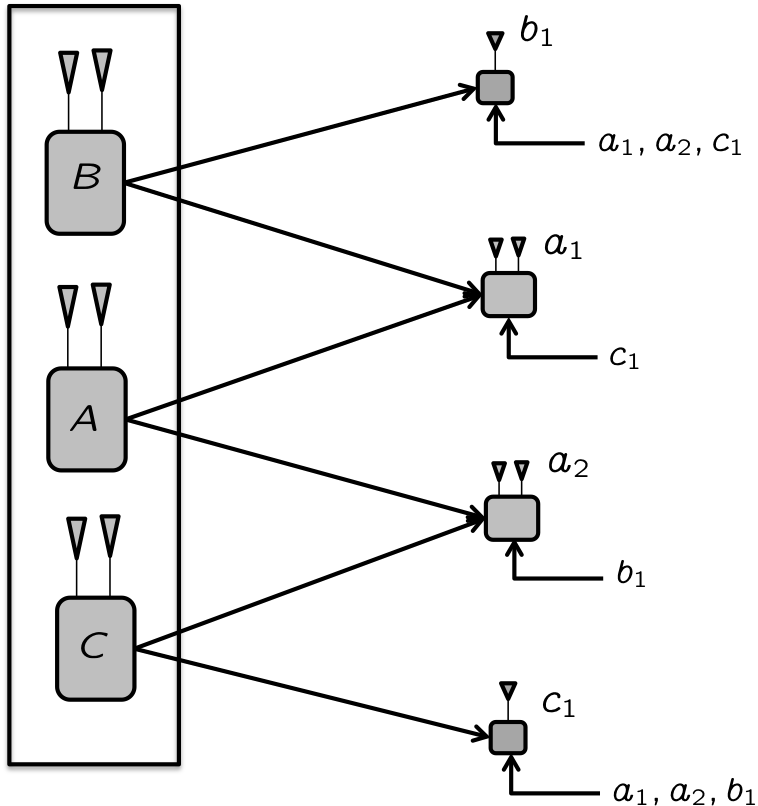}
\caption{\small\it Gaussian Index Coding problem. Each of the 4 messages achieves 1 DoF simultaneously due to message alignment.}
\label{fig:3cellMIMOindex}
\end{figure}

In going from the CB problem to the GIC problem, two key steps were involved. Providing the messages as genie side information from each transmitter to disconnected receivers, and allowing transmitters to cooperate. Suppose we perform both these operations for the heterogeneous MIMO network of Fig. \ref{fig:3cellMIMO}, producing the Gaussian index coding problem shown in Fig. \ref{fig:3cellMIMOindex}. Note that we have not made the system fully connected or used the same marginals property or provided any CSIT, because these operations are non-essential to the present discussion. What we want to point out is that  transmitter cooperation and receiver cognition (messages provided as side information to receivers) open the door to \emph{message alignments} in the (Gaussian) index coding setting that are not available in the original CB problem. To see this, let us consider the DoF for the MIMO BC with cognitive receivers and no CSIT, shown in Fig. \ref{fig:3cellMIMOindex}. It turns out that each of the 4 messages can achieve 1 DoF simultaneously in this problem due to \emph{message} alignment (essentially, network coding). Specifically, at the transmitter, let us XOR the bits for messages $a_1, c_1$ to create a new message $W_{ac}=a_1 + c_1$, and similarly let us XOR the bits for messages $a_2, b_1$ to create a new message $W_{ab}=a_2+b_1$. We only transmit messages $W_{ab}, W_{ac}$. This can be done, e.g., by sending $W_{ab}$ from the first antenna of transmitters $A, B, C$ and $W_{ac}$ from the second antenna of transmitters $A, B, C$. It is easy to verify that all receivers are able to decode their desired messages. Receiver $c_1$ already know one of the messages $(W_{ab})$, so with a single antenna it is able to decode message $W_{ac}$, and also knowing $a_1$, it is able to recover the desired message $c_1$ from $W_{ac}$. By symmetry receiver $b_1$ is able to recover message $b_1$ as well. Receivers $a_1, a_2$ have two antennas each, so each of them can resolve and decode both messages $W_{ab}, W_{ac}$, and having prior knowledge of $c_1, b_1$, can recover their desired messages $a_1, a_2$ from $W_{ac}, W_{ab}$, respectively. 

The message alignments make it possible to achieve a total of 4 DoF in the Gaussian index coding problem of Fig. \ref{fig:3cellMIMOindex}. The message alignment solution in turn is possible mainly because of transmitter cooperation and receiver cognition, both of which are essential to (Gaussian) index coding. However, neither transmitter cooperation nor receiver cognition is available in the original cellular blind interference alignment problem of Fig. \ref{fig:3cellMIMO}, making message alignment solutions infeasible. On the other hand, if we consider only linear beamforming solutions, which can translate between GIC and CB problems, it seems difficult to exploit either message cognition or transmitter cooperation to improve upon the solution of the CB problem, i.e., to achieve more than $4/3$ DoF for cell A, while achieving $1$ DoF for each of cells $B, C$ simultaneously. Thus, the wireless index coding problem is like the Gaussian index coding problem, restricted to linear beamforming solutions. If in addition, sufficient coherence cannot be assumed, then there are also further restrictions on the beamforming solutions,  such as the range of the support of the beamforming vectors. Thus, both the similarities and differences between index coding and cellular blind interference alignment are quite  intriguing as well as significant.

To complete the discussion of the example of Fig. \ref{fig:3cellMIMO}, we point out that with perfect CSIT, this network has 4 DoF. This can be seen as follows. Messages $a_1, b_1$ cannot  have a total of more than 2 DoF, because of the MIMO interference channel DoF outer bound of \cite{Jafar_Fakhereddin}. By symmetry, messages $a_2, c_1$ cannot have more than 2 DoF either, which limits the total DoF to 4. Achievability of 4 DoF is also easy because after removing the interference, receivers $a_1, c_1$ are left with one antenna each, so that the resulting MISO BC in cell $A$ can achieve 2 DoF by zero forcing. 

The reciprocal network of Fig. \ref{fig:3cellMIMO}, obtained by switching the roles of the transmitters and receivers for each message, is easily verified to have 4 DoF without CSIT.   

Finally, if both CSIT and transmitter cooperation are allowed, or if simply receiver cooperation is allowed, then the network has 6 DoF, which is simply the rank of the overall channel matrix.

\section{Conclusion}
We explore the cellular blind interference alignment problem, i.e., the degrees of freedom of wireless networks with no CSIT. Somewhat surprisingly this highly relevant problem seems to have received little attention, perhaps because of the conventional wisdom that such pessimistic assumptions --- no CSIT, statistically i.i.d. fading, single antenna users, homogeneous coherence intervals, no reconfigurable antennas --- can only lead to degenerate solutions, such as the traditionally well-studied orthogonal solutions. In this work, our goal is to highlight the challenging nature of the problem, surprising solutions that arise in this setting even when orthogonal (aligned frequency reuse), its intimate relationship to the index coding problem, a problem already known to be hard, and how the restrictions of wireless settings add further challenges and lead to a truly wireless index coding problem, and finally how interference diversity, always available in a wireless network but so far ignored in DoF studies, can create opportunities for robust blind interference alignment solutions. We believe this work has only scratched the surface of this most interesting research direction, and that further research will continue to reveal additional surprising and fundamental aspects of the cellular blind interference alignment problem.

\onehalfspacing 

\bibliographystyle{IEEEtran}
\bibliography{Thesis}

\begin{thebibliography}{10}
\providecommand{\url}[1]{#1}
\csname url@samestyle\endcsname
\providecommand{\newblock}{\relax}
\providecommand{\bibinfo}[2]{#2}
\providecommand{\BIBentrySTDinterwordspacing}{\spaceskip=0pt\relax}
\providecommand{\BIBentryALTinterwordstretchfactor}{4}
\providecommand{\BIBentryALTinterwordspacing}{\spaceskip=\fontdimen2\font plus
\BIBentryALTinterwordstretchfactor\fontdimen3\font minus
  \fontdimen4\font\relax}
\providecommand{\BIBforeignlanguage}[2]{{%
\expandafter\ifx\csname l@#1\endcsname\relax
\typeout{** WARNING: IEEEtran.bst: No hyphenation pattern has been}%
\typeout{** loaded for the language `#1'. Using the pattern for}%
\typeout{** the default language instead.}%
\else
\language=\csname l@#1\endcsname
\fi
#2}}
\providecommand{\BIBdecl}{\relax}
\BIBdecl

\bibitem{Cadambe_Jafar_X}
V.~Cadambe and S.~Jafar, ``Interference alignment and the degrees of freedom of
  wireless {X} networks,'' \emph{IEEE Trans. on Information Theory}, no.~9, pp.
  3893--3908, Sep 2009.

\bibitem{Jafar_FnT}
S.~Jafar, ``Interference alignment: A new look at signal dimensions in a
  communication network,'' in \emph{Foundations and Trends in Communication and
  Information Theory}, vol.~7, no.~1, 2011, pp. 1--136.

\bibitem{Cadambe_Jafar_int}
V.~Cadambe and S.~Jafar, ``Interference alignment and the degrees of freedom of
  the {K} user interference channel,'' \emph{IEEE Transactions on Information
  Theory}, vol.~54, no.~8, pp. 3425--3441, Aug. 2008.

\bibitem{Maddah_Tse}
M.~A. Maddah-Ali and D.~Tse, ``Completely stale transmitter channel state
  information is still very useful,'' \emph{48th Annual Allerton Conference on
  Communication, Control and Computing}, 2010.

\bibitem{Maleki_Jafar_Shamai}
H.~Maleki, S.~A. Jafar, and S.~Shamai, ``{Retrospective Interference Alignment
  over Interference Networks},'' \emph{IEEE Journal of Selected Topics in
  Signal Processing}, 2012.

\bibitem{Lapidoth_Shamai_Wigger_BC}
A.~Lapidoth, S.~Shamai, and M.~Wigger, ``On the capacity of fading {MIMO}
  broadcast channels with imperfect transmitter side-information,'' in
  \emph{Proceedings of 43rd Annual Allerton Conference on Communications,
  Control and Computing}, Sep. 28-30, 2005.

\bibitem{Weingarten_Shamai_Kramer}
H.~Weingarten, S.~Shamai, and G.~Kramer, ``On the compound {MIMO} broadcast
  channel,'' in \emph{Proceedings of Annual Information Theory and Applications
  Workshop UCSD}, Jan 2007.

\bibitem{Huang_Jafar_Shamai_Vishwanath}
C.~Huang, S.~A. Jafar, S.~Shamai, and S.~Vishwanath, ``{On Degrees of Freedom
  Region of MIMO Networks without Channel State Information at Transmitters},''
  \emph{IEEE Transactions on Information Theory}, no.~2, pp. 849--857, Feb.
  2012.

\bibitem{Zhu_Guo_MIMOIC}
Y.~Zhu and D.~Guo, ``The degrees of freedom of isotropic {MIMO} interference
  channels without state information at the transmitters,'' \emph{IEEE
  Transactions on Information Theory}, vol.~58, no.~1, pp. 341--352, 2012.

\bibitem{Guo_isotropic}
\BIBentryALTinterwordspacing
------, ``Isotropic {MIMO} interference channels without { CSIT}: The loss of
  degrees of freedom,'' \emph{47th Annual Allerton Conference on Communication,
  Control, and Computing}, vol. abs/0910.2961, 2009. [Online]. Available:
  \url{http://arxiv.org/abs/0910.2961}
\BIBentrySTDinterwordspacing

\bibitem{Varanasi_noCSIT}
\BIBentryALTinterwordspacing
C.~S. Vaze and M.~K. Varanasi, ``The degrees of freedom regions of {MIMO}
  broadcast, interference, and cognitive radio channels with no {CSIT},''
  \emph{CoRR}, vol. abs/0909.5424, 2009. [Online]. Available:
  \url{http://arxiv.org/abs/0909.5424}
\BIBentrySTDinterwordspacing

\bibitem{Gou_Jafar_Wang}
T.~Gou, S.~Jafar, and C.~Wang, ``On the degrees of freedom of finite state
  compound wireless networks,'' \emph{IEEE Transactions on Information Theory},
  vol.~57, no.~6, pp. 3268--3308, June 2011.

\bibitem{Jafar_corr}
S.~A. Jafar, ``{Blind Interference Alignment},'' \emph{IEEE Journal of Selected
  Topics in Signal Processing}, 2012.

\bibitem{Wang_Gou_Jafar}
T.~Gou, C.~Wang, and S.~A. Jafar, ``Aiming perfectly in the dark - blind
  interference alignment through staggered antenna switching,'' \emph{IEEE
  Trans. on Signal Processing}, vol.~59, pp. 2734--2744, June 2011.

\bibitem{Wang_Gou_Jafar_MIMO}
C.~Wang, T.~Gou, and S.~A. Jafar, ``Interference alignment through staggered
  antenna switching for {MIMO} {BC} with no {CSIT},'' in \emph{Asilomar
  Conference on Signals, Systems and Computers}, Nov. 2010.

\bibitem{Birk_Kol}
Y.~Birk and T.~Kol, ``{Informed-source coding-on-demand (ISCOD) over broadcast
  channels},'' in \emph{Proceedings of the Seventeenth Annual Joint Conference
  of the IEEE Computer and Communications Societies, IEEE INFOCOM'98}, vol.~3,
  1998, pp. 1257--1264.

\bibitem{Koetter_Effros_Medard}
R.~Koetter, M.~Effros, and M.~Medard, ``{On a theory of network equivalence},''
  \emph{IEEE Information Theory Workshop on Networking and Information Theory},
  pp. 326--330, 2009.

\bibitem{Blasiak_Kleinberg_Lubetzky_2011}
\BIBentryALTinterwordspacing
A.~Blasiak, R.~Kleinberg, and E.~Lubetzky, ``Lexicographic products and the
  power of non-linear network coding,'' \emph{CoRR}, Aug. 2011. [Online].
  Available: \url{http://arxiv.org/abs/1108.2489}
\BIBentrySTDinterwordspacing

\bibitem{Rouayheb_Sprintson_Georghiades}
S.~Rouayheb, A.~Sprintson, and C.~Georghiades, ``{On the Index Coding Problem
  and Its Relation to Network Coding and Matroid Theory},'' \emph{IEEE
  Transactions on Information Theory}, vol.~56, no.~7, pp. 3187--3195, July
  2010.

\bibitem{Blasiak_Kleinberg_Lubetzky_2010}
\BIBentryALTinterwordspacing
A.~Blasiak, R.~Kleinberg, and E.~Lubetzky, ``Index coding via linear
  programming,'' \emph{CoRR}, April 2010. [Online]. Available:
  \url{http://arxiv.org/abs/1004.1379}
\BIBentrySTDinterwordspacing

\bibitem{Maleki_Cadambe_Jafar}
H.~Maleki, V.~Cadambe, and S.~Jafar, ``Index coding -- an interference
  alignment perspective,'' \emph{submitted to ISIT 2012}, 2012.

\bibitem{Jafar_Fakhereddin}
S.~Jafar and M.~Fakhereddin, ``Degrees of freedom for the {MIMO} interference
  channel,'' \emph{IEEE Transactions on Information Theory}, vol.~53, no.~7,
  pp. 2637--2642, July 2007.

\end{thebibliography}

\end{document}